\documentclass[twocolumn]{aastex62}
\hypersetup{breaklinks}
\usepackage{comment}

\newcommand{\teff}{\ensuremath{T_{\mbox{\scriptsize eff}}}}

\newcommand{\msun}{\ensuremath{\mbox{M}_{\odot}}}

\newcommand{\prot}{\ensuremath{P_{\mbox{\scriptsize rot}}}}

\newcommand{\soren}{S\o ren Meibom}
\newcommand{\degree}{\ensuremath{^\circ}}
\newcommand{\mas}{\ensuremath{\mbox{mas yr}^{-1}}}

\newcommand{\gbr}{\ensuremath{(G_{\rm BP} - G_{\rm RP})}}

\accepted{May 16, 2019}
\submitjournal{\textit{The Astrophysical Journal}}

\shorttitle{Stellar Spin-Down in NGC 6811 with \textit{Gaia} and \textit{Kepler}}
\shortauthors{Curtis, Ag\"{u}eros,  Douglas, \& Meibom}

\begin{document}

\title{{\sc A Temporary Epoch of Stalled Spin-Down for Low-Mass Stars:\\Insights from NGC 6811 with  \textit{Gaia} and \textit{Kepler}}}

\newcommand{\cfa}{Center for Astrophysics $|$ Harvard \& Smithsonian, 60 Garden Street, Cambridge, MA 02138, USA} 
\newcommand{\columbia}{Department of Astronomy, Columbia University, 550 West 120th Street, New York, NY 10027, USA}

\correspondingauthor{Jason Lee Curtis}
\email{jasoncurtis.astro@gmail.com}
\author[0000-0002-2792-134X]{Jason Lee Curtis}
\altaffiliation{NSF Astronomy and Astrophysics Postdoctoral Fellow}
\affiliation{\columbia}

\author[0000-0001-7077-3664]{Marcel A.~Ag\"{u}eros}
\affiliation{\columbia}

\author[0000-0001-7371-2832]{Stephanie T.~Douglas}
\altaffiliation{NSF Astronomy and Astrophysics Postdoctoral Fellow}
\affiliation{\cfa}

\author{\soren}
\affiliation{\cfa}


\begin{abstract}
Stellar rotation was proposed as a potential age diagnostic that is precise, simple, and applicable to a broad range of low-mass stars ($\leq$1~\msun). Unfortunately, rotation period (\prot) measurements of low-mass members of open clusters have undermined the idea that stars spin down with a common age dependence (i.e., \prot\ $\propto \sqrt{\rm age}$): K dwarfs appear to spin down more slowly than F and G dwarfs. \citet{Agueros2018} interpreted data for the $\approx$1.4-Gyr-old cluster NGC 752 differently, proposing that after having converged onto a slow-rotating sequence in their first 600--700~Myr (by the age of Praesepe), K dwarf \prot\ stall on that sequence for an extended period of time. 
We use data from \textit{Gaia} DR2 to identify likely single-star members of the $\approx$1-Gyr-old cluster NGC 6811 with \textit{Kepler} light curves. We measure \prot\ for 171 members, more than doubling the sample relative to the existing catalog and extending the mass limit from 
$\approx$0.8 to $\approx$0.6~\msun.
We then apply a gyrochronology formula calibrated with Praesepe and the Sun to 27 single G dwarfs in NGC 6811 to derive a precise gyrochronological age for the cluster of 1.04$\pm$0.07~Gyr. However, when our new low-mass rotators are included, NGC 6811's color--\prot\ sequence deviates away from the naive 1~Gyr projection down to $\teff \approx 4295$ K (K5V, 0.7~\msun), where it clearly overlaps with Praesepe's. 
Combining these data with \prot\ for other clusters, we conclude that
the assumption that mass and age are separable dependencies is invalid. Furthermore, the cluster data show definitively that stars experience a temporary
epoch of reduced braking efficiency where \prot\ stall, and that the duration of this epoch lasts longer for lower-mass stars.
\end{abstract}





\section{Introduction} \label{s:intro}
Sun-like stars change very little over their main-sequence lifetimes, making their ages one of their most challenging properties to determine.
However, knowing stellar ages, especially for low-mass stars ($\leq$1~\msun), is essential in this era of precision astrophysics. On the Galactic scale, our current inability to provide confident ages for these stars limits our understanding of the Milky Way's star-formation history and chemical enrichment. On the planetary-system scale, it negatively impacts the development of theories for planet formation and evolution. 

Forty-seven years ago, \citet{skumanich1972} compared rotation for solar-mass members of young nearby clusters (the Pleiades, Ursa Major, and the Hyades) to the Sun, and noted that stars appeared to spin down according to the 
square-root of age,
which has since become known as the Skumanich Law. \cite{Barnes2003} built on this work to propose the use of rotation periods (\prot) as a clock, which he termed gyrochronology.
A reliable rotation--age relation would be a boon to the study of these stars, because other techniques for obtaining their ages generally do not work \citep{Soderblom2010}. 
For example, for K and M stars, relying on evolution off of the zero-age main-sequence and the subsequent increase in luminosity or decrease in surface gravity cannot lead to meaningful constraints on ages, as these parameters are practically unchanged over the age of the Universe.

Unfortunately, observational efforts to constrain the rotation--age relation, based largely on observations of open clusters, have made our hopes for a simple relation fade. 
For example, when comparing the color--\prot\ distributions for M35 (150~Myr) and M34 (220~Myr) to that for the Hyades (727~Myr\footnote{\citet{Douglas2019} calculated a differential gyrochronology age of 727~Myr for the Hyades relative to Praesepe, which we fixed to 670~Myr based on the median of several literature isochrone ages.}), 
\citet{MeibomM35, MeibomM34} found that while the F and G dwarfs had spun down following the Skumanich Law, 
K-type Hyads appeared to rotate too rapidly relative to their younger counterparts in M35 and M34, after projecting each sample to a common age using $\prot \propto t^{0.5}$. 
\citet{Meibom2011} noted a similar behavior for the early K  dwarfs in NGC~6811 (1~Gyr). 
\citet{Cargile2014} reached a similar conclusion upon comparing rotation 
data for Blanco~1 \citep[132~Myr;][]{Cargile2010} to these same clusters.
These authors suggested that Sun-like FGK dwarfs spin down continuously, but with time dependencies that differ according to mass.

\citet{barnes2007, Barnes2003} derived gyrochronology relations that
decoupled the mass and age dependence of stellar spin-down. 
Any mass dependence could then be determined from the \prot\ sequences observed in 
young, nearby clusters, and the age index $n$, also known as the braking index, for the $t^n$ power law could be fitted for by comparing those sequences to the Sun's \prot\ and age. 

However, while re-tuning the coefficients for the \citet{barnes2007} gyrochronology equation, 
\citet{Angus2015} were forced to discard data for Praesepe (670~Myr) and NGC 6811. 
In addition, \citet{Agueros2018} analyzed light curves from the Palomar Transient Factory \citep[PTF;][]{nick2009,rau2009} for the $\approx$1.4-Gyr-old\footnote{\citet{Agueros2018} inferred an age of 1.34 Gyr for NGC 752; \citet{Twarog6819} found 1.45 Gyr. We average the two results and round it to 1.4 Gyr.} cluster NGC 752 
and found that the cluster's early K dwarfs had barely slowed relative to Praesepe's, while the late K and early M dwarfs had not spun down at all, despite being about twice as old.

Here, we 
re-examine rotation in the 1 Gyr cluster NGC 6811. We use high-precision astrometry and photometry from \textit{Gaia} to identify cluster members with \textit{Kepler} light curves, which extends the 1-Gyr rotator sample from $\approx$0.8~\msun\ \citep{Meibom2011} down to $\approx$0.6~\msun\ (Section~\ref{s:rot}). Next, we compare our expanded sample to the rotation data for the younger cluster Praesepe, 
and show that while the F and G dwarfs have spun down as expected, the K dwarfs have not slowed at all in the intervening $\approx$350~Myr. Finally, we demonstrate that this cannot be due to K dwarfs simply spinning down more slowly than F/G dwarfs, but instead must be caused by a temporary period of stalled braking (Section~\ref{s:dis}). We conclude in Section~\ref{s:concl}.

\section{New Rotators in NGC 6811} \label{s:rot}
Below, we review the properties of NGC 6811, first observed by John Herschel in 1829 \citep{FirstR147ref}. We then expand the cluster's membership with data from the second \textit{Gaia} data release \citep[DR2;][]{GaiaDR2} and determine the properties of these stars. We also discuss the rotation results from \citet{Meibom2011} before measuring \prot\ for those stars with \textit{Kepler} \citep{Kepler2005} light curves and producing a new color--\prot\ distribution for the cluster.

\subsection{The age, metallicity, and reddening of NGC 6811}
\citet{Sandquist2016} presented a thorough analysis of NGC 6811, including the characterization of a detached partially eclipsing binary (EB), measurements of asteroseismic parameters for helium-burning giants, study of the pulsating stars at the main-sequence turnoff, and an analysis of its color--magnitude diagram (CMD). While there is some tension between the age solutions for the two EB components, most of the data support an age for the cluster of $1 \pm 0.05$ Gyr \citep{Sandquist2016}, in agreement with the \citet{Janes6811} $UBVRI$ photometric analysis.

\citet{Sandquist2016} also summarized the spectroscopic metallicity measurements in the literature from various sources, which agree on an approximately solar metallicity with values ranging from $-0.02$ to $+0.04$ dex. 
Finally, \citet{Sandquist2016}  found an interstellar reddening value of $E(B - V) = 0.07 \pm 0.02$ (corresponding to $A_V = 0.22$), while various literature values range from $E(B - V) = 0.05$ to 0.14 ($A_V = 0.16$ to 0.43).
This is consistent with the 3D dust map value built with 
Pan-STARRS 1 and 2MASS photometry, which 
estimates $A_V = 0.25 \pm 0.06$ at 1.15~kpc \citep{Green2018}.\footnote{\url{http://argonaut.skymaps.info}}
In Section~\ref{s:red}, we refine the reddening/extinction value
by fitting a gyrochronology model to the color--period distribution
and find $A_V = 0.15$.
\subsection{Identifying NGC 6811 members with \textit{Gaia}}\label{new_id}

\begin{figure*}[ht!]\begin{center}
\includegraphics[width=3.4in]{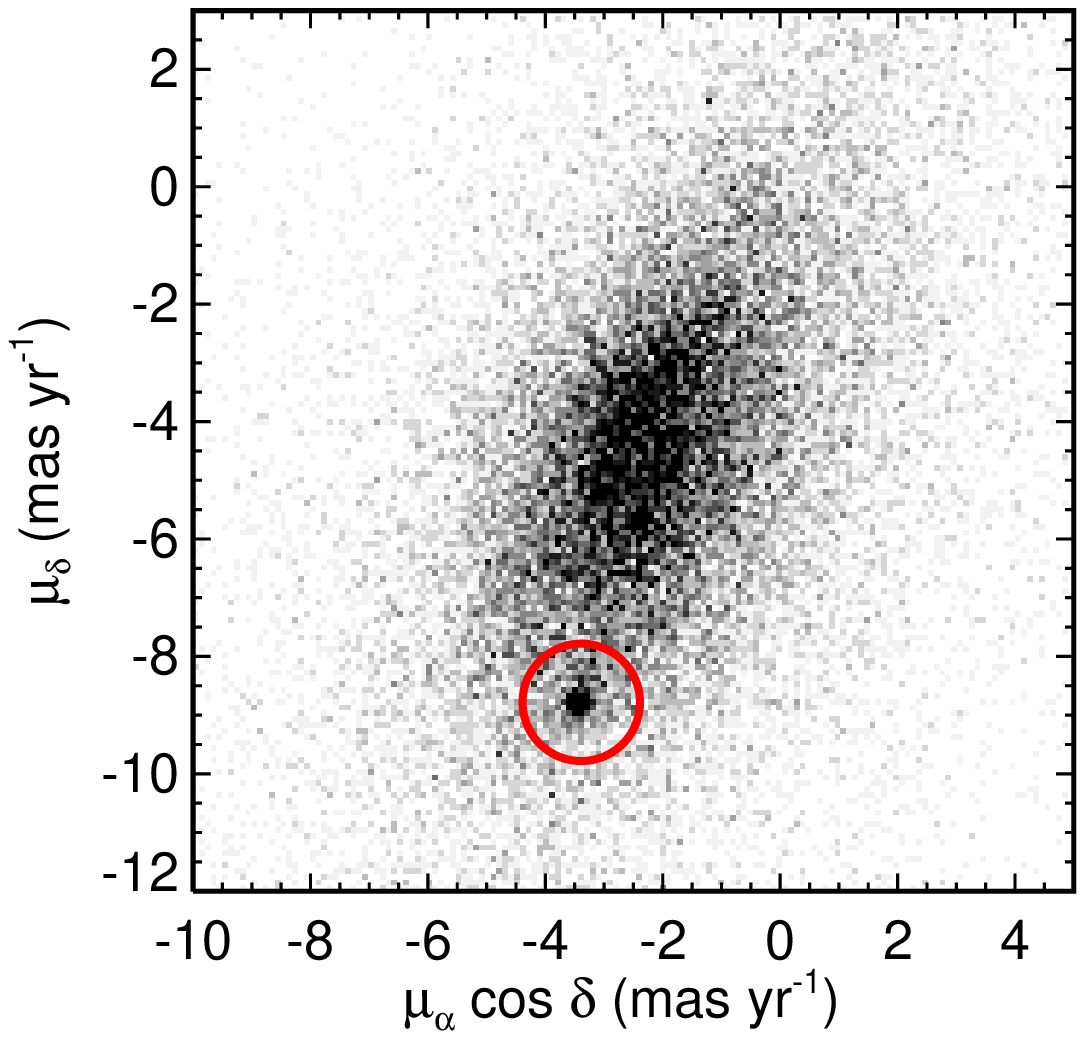} 
\includegraphics[width=3.4in]{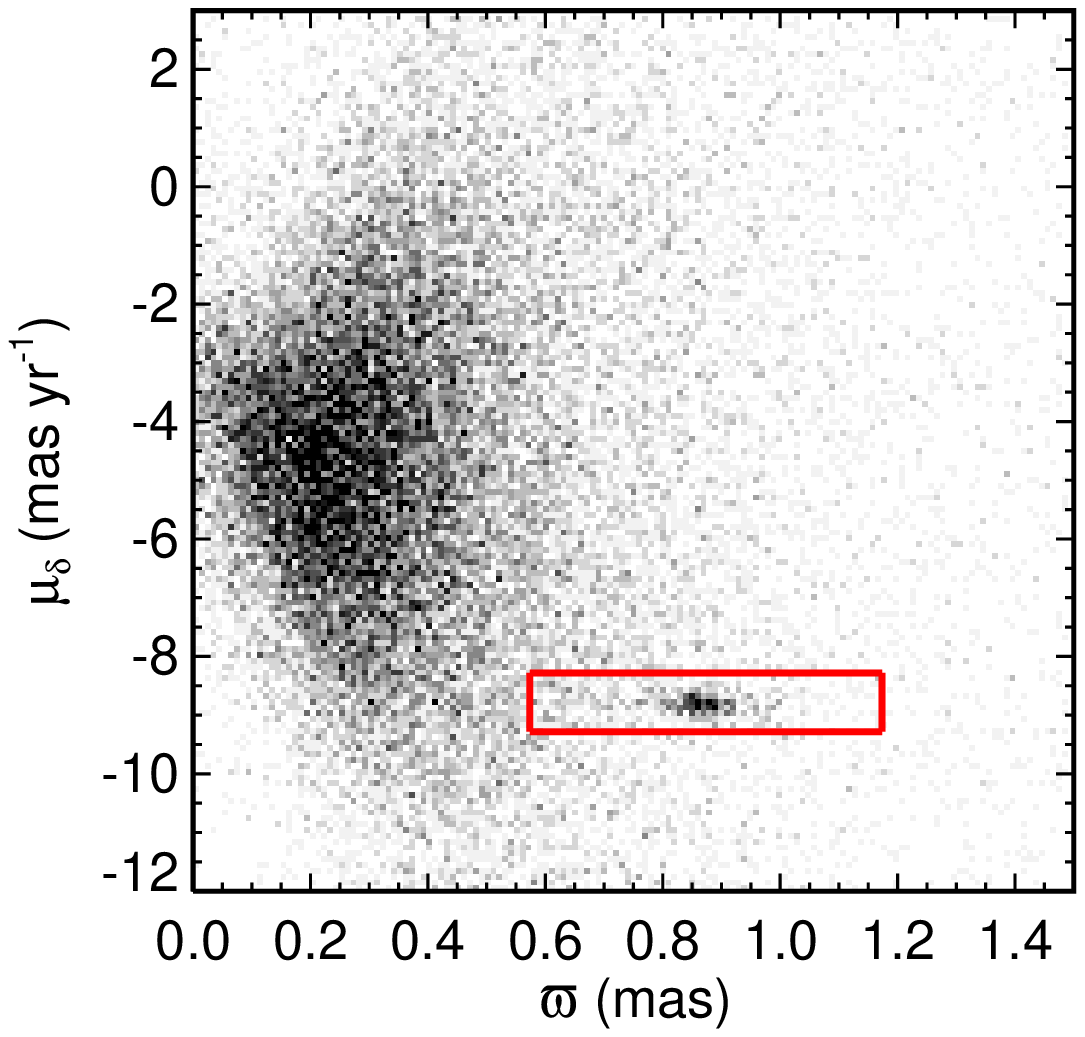} 
\includegraphics[trim=1.1cm 0cm 0.3cm 0cm, clip=True, width=2.34in]{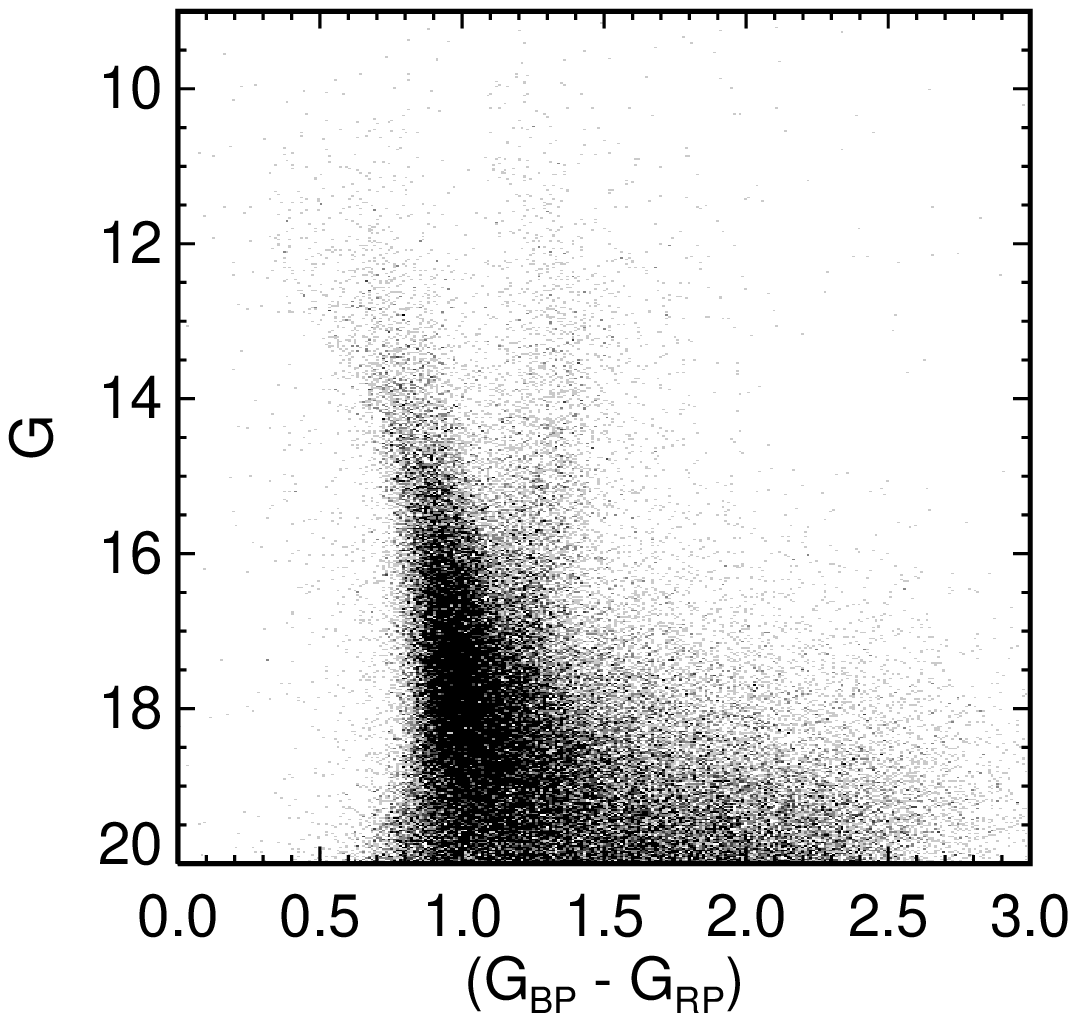} 
\includegraphics[trim=1.1cm 0cm 0.3cm 0cm, clip=True, width=2.34in]{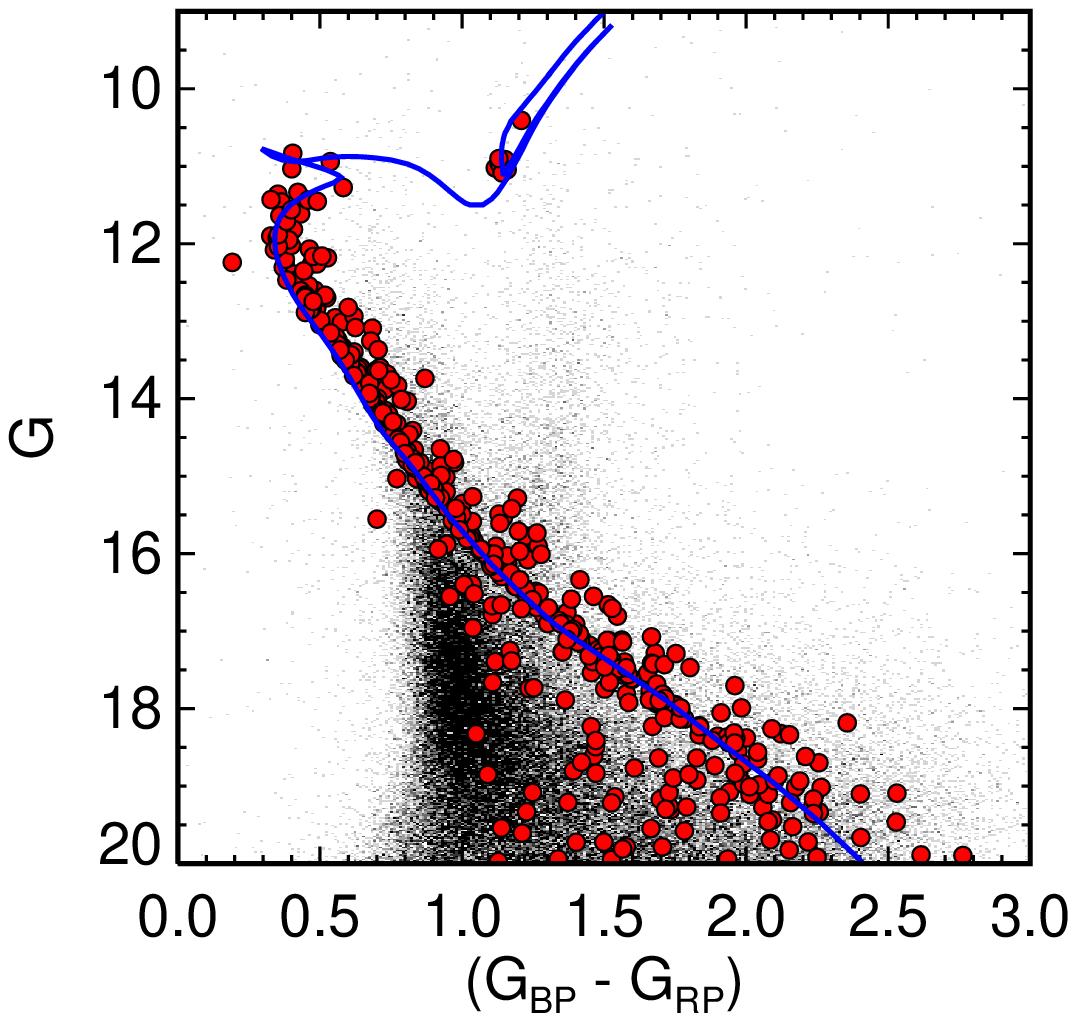} 
\includegraphics[trim=1.1cm 0cm 0.3cm 0cm, clip=True, width=2.34in]{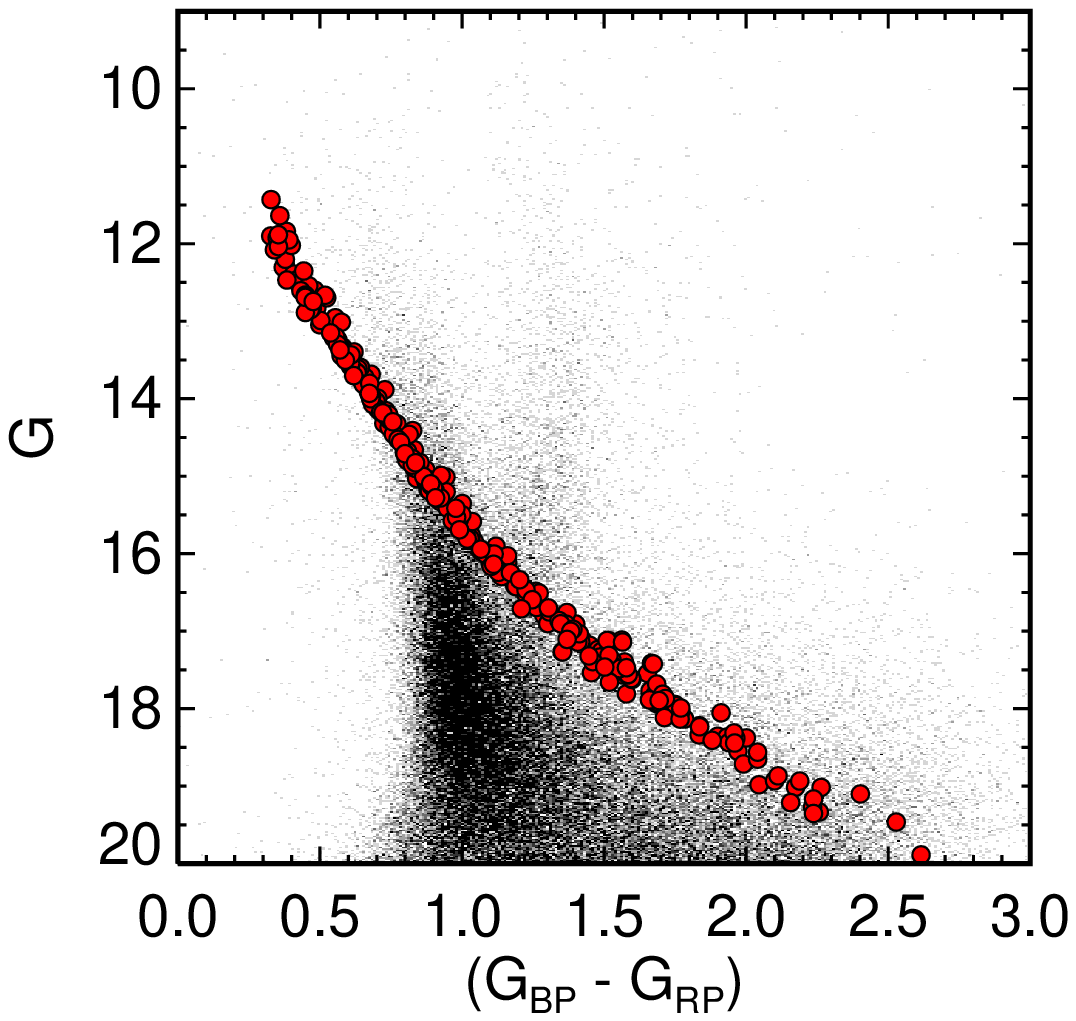} 
  \caption{Selection of single-star members of NGC 6811 with \textit{Gaia} DR2. 
  \textit{Top left---}Proper motions of stars within 1\degree\ of NGC~6811 and $G < 18$ mag (we considered stars with $G < 20$; we adjust this threshold for clarity in this figure); the red circle has a radius of 1 \mas, which is double the value we adopted for membership criteria (0.5 \mas) to make this panel easier to view.
  \textit{Top right---}Parallax versus declination proper motion for the same stars.
  The red box shows our selection criteria
  (0.3 mas in parallax, and again 0.5 \mas\ in proper motion). 
  \textit{Bottom left---}The CMD for 
  the $\approx$87,000 stars in the field.  
  \textit{Bottom middle---}The 485 stars satisfying our astrometric criteria are highlighted in red. A 1~Gyr PARSEC model with solar composition, 
  $A_V = 0.15$, and $(m - M) = 10.2$ is overlaid in blue.
  \textit{Bottom right---}The subset of 322 astrometric (unevolved) members that follow the cluster's single-star sequence are  highlighted in red.
   \label{f:mem}}
\end{center}\end{figure*}

The \cite{Meibom2011} study of rotation in NGC 6811 relied on a sample vetted with radial velocities (RVs) to clean the {\it Kepler} $(g - r)_{\rm KIC}$ vs.~$g_{\rm KIC}$ CMD.  \textit{Gaia} recently delivered high-precision astrometry (positions, proper motions, parallaxes) and photometry ($G$, $G_{\rm BP}$, $G_{\rm RP}$) for $>$$1.3\times10^9$ stars \citep[][]{GaiaDR2}. These data greatly simplify the process of identifying single-star members of open clusters \citep[e.g.,][]{DR2HRD}.

We determine the cluster's DR2 proper motion ($\mu$) and parallax ($\varpi$) from the 71 members identified by \citet{Meibom2011}: 
$\mu_\alpha \cos \delta = -3.39$~\mas, 
$\mu_\delta = -8.78$~\mas, and
$\varpi = 0.87$~mas.
There are 86,853 stars within 1\degree\ of NGC 6811's center
(a 19~pc search radius at $d = $1096~pc)\footnote{$(\alpha, \delta) = 19^{\rm h}37^{\rm m}12^{\rm s}$, $+46\degree23'15''$.} in DR2 with 
$G < 20$ mag and usable 
astrometry and photometry, but
only 485 satisfy our simple astrometric criteria: $\mu$ within 0.5~\mas\ and $\varpi$ within 0.3~mas of the median for these 71 stars (see top row of Figure \ref{f:mem}). 

After applying our astrometric cuts, we overplot the candidate cluster members on the {\it Gaia} CMD for the region around NGC 6811.
(see the bottom row of Figure \ref{f:mem}). The cluster's CMD is visually well-fit with a PARSEC isochrone model \citep{parsec, Chen2014} for an age of  
1~Gyr, solar composition, $A_V = 0.15$,
and $(m - M) = 10.2$, as shown in the bottom middle panel of Figure \ref{f:mem}. 

\begin{figure*}\begin{center}
\includegraphics[width=3.4in]{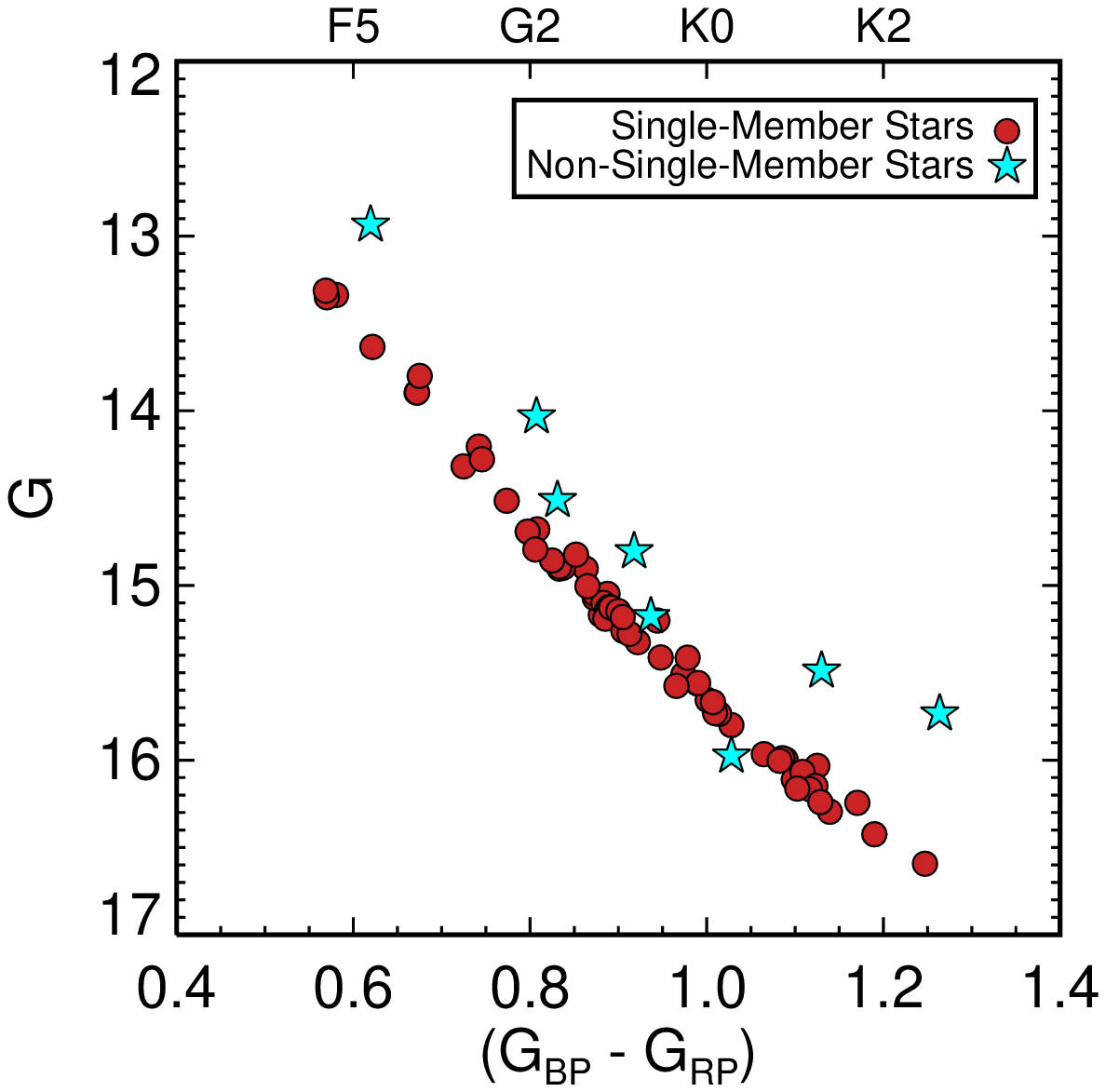} 
\includegraphics[width=3.4in]{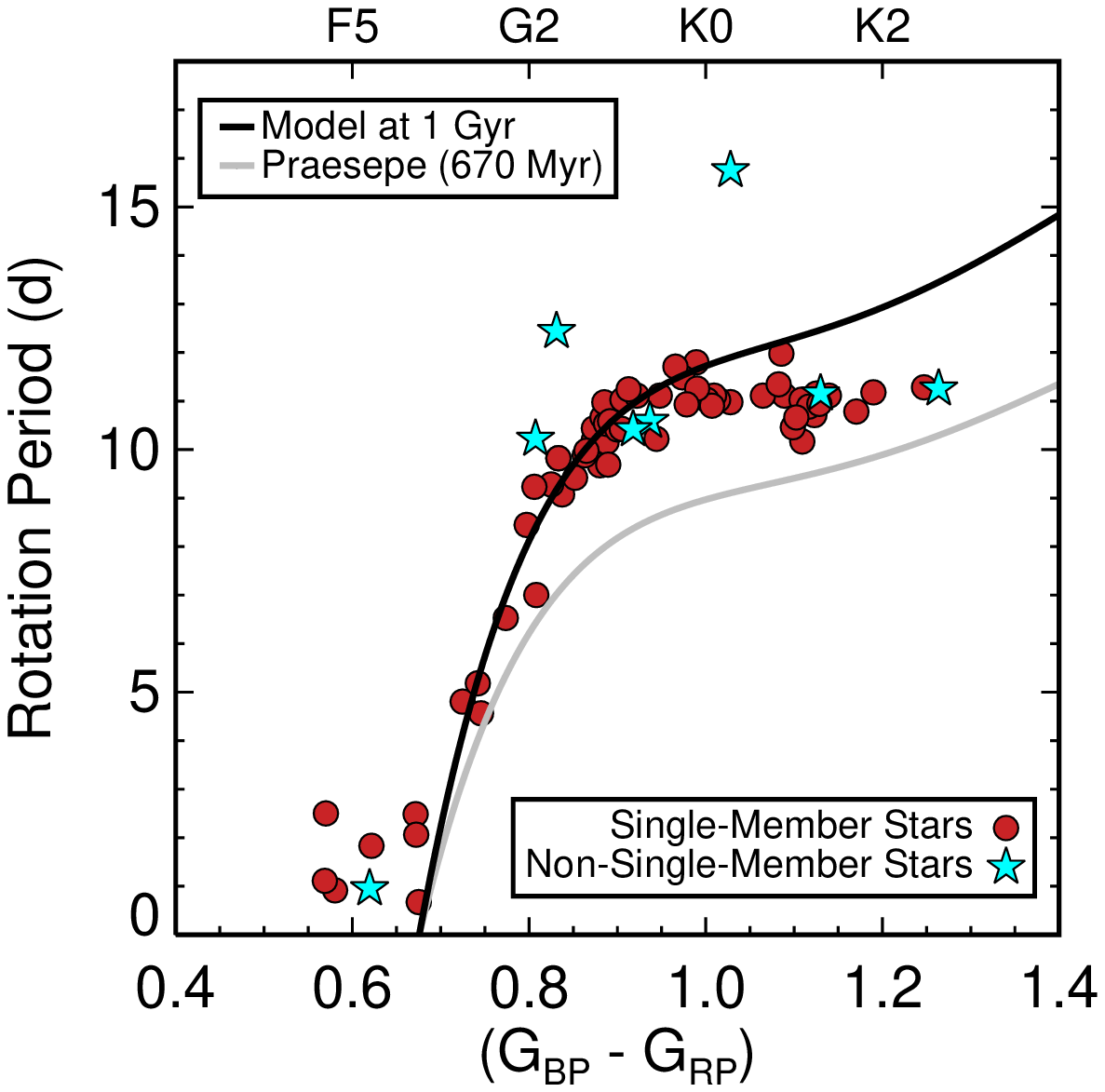} 
  \caption{CMD (\textit{left}) and color--period diagram (\textit{right}) for the 71 stars in the \citet{Meibom2011} rotator sample for NGC 6811.
  We identified eight photometric and proper motion outliers (cyan stars), two of which are also outliers in the color--period diagram.
  A 1 Gyr gyrochronology model (upper black line), calibrated from fitting the Praesepe color--period sequence (lower gray line) 
  and tuned with the Sun, is also included.
  This model, with braking index $n = 0.62$, fits the F and early-G dwarfs, but diverges from the later-G and K dwarfs. 
  \label{f:mei}}
\end{center}\end{figure*}

We then use the Hyades main-sequence from \citet{DR2HRD} to define an empirical single-star sequence, which we fit with a cubic basis spline to predict $M_G$ from \gbr. Next, we apply the Hyades model to the NGC 6811 astrometric candidates using the NGC 6811 extinction and distance modulus, and filter out stars that are offset by more than 0.5~mag in $M_G$ from the model. 

We fit a cubic basis spline to the stars remaining in the NGC 6811 sample to define the cluster's single-star sequence more accurately, and then re-extract the single stars with a stricter cut of 0.25~mag. In this manner, we identify 322 likely single members along the main-sequence of NGC 6811 (see bottom right panel of Figure~\ref{f:mem}). While this is not a complete census of the cluster's membership, this new catalog is all that is required for this work, which focuses on the rotational behavior of single stars.\footnote{While characterizing the binary population is critical in any cluster, in this context binaries are contaminants. Binary companions may exert tidal or other physical effects on the primary star \citep[e.g.,][]{meibom2005, meibom2007,douglas2016,douglas2017}, causing us to locate stars incorrectly in the color--\prot\ plane, and leading us to misidentify trends or transitions in the period distribution.} 

\subsection{Determining stellar properties} 
While the DR2 stellar properties pipeline \citep[Apsis;][]{apsis2013,DR2prop} produced effective temperatures (\teff) from the DR2 photometry for $1.61\times10^8$ stars with $G<17$~mag and $3000 < \teff < 10,000$~K, these notably do not incorporate the effects of reddening. We therefore generate a color--\teff\ relation
using three separate catalogs of precisely characterized nearby stars: the \citet{Brewer2016} sample of FGK stars observed by the California Planet Search with Keck/HIRES,
the \citet{Boyajian2012} sample of K and M dwarfs with interferometric radii and bolometric fluxes, and the \citet{Mann2015} M dwarfs characterized with optical and near-infrared spectroscopy. Deriving our own empirical relation also allows us to correct for NGC 6811's metallicity.\footnote{Throughout this work, we use the DR2 extinction coefficients for the \citet{DR2phot2} passbands provided by the PARSEC isochrone web page (\url{http://stev.oapd.inaf.it/cgi-bin/cmd_3.0}) for a G2V star using the \citet{Cardelli1989} extinction law with $R_V = 3.1$, where $E\gbr / A_V \approx 0.416$ and $A_G / A_V \approx 0.859$, and ignore their dependence on spectral type.}

The temperatures of stars hotter than 4100 K in our benchmark sample are consistent with the DR2 values to within 70~K (root-mean-square error), with most of the remaining scatter due to the \gbr\ color's dependence on metallicity along with \teff\ \citep[see figure 2 of][]{Morris2018}. However, the Apsis \teff\ values diverge from those in our sample for $\teff < 4100$~K. 

We also translate our photometric \teff\ values to spectral types (SpT) and stellar masses ($M_\star$) using the \citet{kraus2007} stellar spectral energy distribution table (their table 5) to aid the reader's interpretation of our results. We interpolate mass from \teff, but  only quote the SpT at the nearest \teff\ entry in the \citet{kraus2007} table, which for our sample includes the following spectral types: F5, F8, G0, G2, G5, G8, K0, K2, K4, K5, K7, M0, and M1. Our \teff, $M_\star$, and SpT values are listed in Table \ref{table} for all the rotators we identify below and use in our analysis.

\subsection{Revisiting the \citet{Meibom2011} results} 
NGC 6811 was observed during the primary \textit{Kepler} mission. \citet{Meibom2011} presented rotation periods measured from light curves for Quarters 1-4 for the 71 members they confirmed by RV monitoring.

Two of these stars appear to rotate too slowly in the color--\prot\ plot for NGC 6811: KICs 9595724 and 9717386. These stars are also outliers in the cluster's DR2 proper motion diagram, and appear too bright in the DR2 CMD, 
which indicates that they are either not members or are binaries. Either way, they should be removed from the gyrochronology calibration sample.

KIC 9594645 also has a discrepant proper motion (by 1 \mas), and appears fainter than the cluster's single-star sequence. Although its \prot\ is consistent with the cluster distribution, we reclassify it as a non-single-member  and remove it from our sample.

We identified five other photometric binaries in the DR2 CMD (KICs 9594100, 9655310, 9471038, 9775381, 9592939). While they are consistent with the cluster color--\prot\ distribution, that distribution is well populated, and we opted to remove these from the sample. 

Figure \ref{f:mei} shows the CMD and color--\prot\ diagram for the \citet{Meibom2011} sample of 71 stars, and highlights the eight stars we identified as non-single-members according to DR2. This leaves 63 stars in the \citet{Meibom2011} sample with masses $0.79 < M_\star < 1.29$ \msun\ (K2V to F5V).

\subsection{Measuring \prot\ with \textit{Kepler} (again)}
We search the Mikulski Archive for Space Telescopes\footnote{\url{https://archive.stsci.edu}} for \textit{Kepler} data for the 278 stars we identify as NGC 6811 members 
with $\gbr \gtrsim 0.57$ 
\citep[the bluest published rotator;][]{Meibom2011}.
We find light curves for 203 stars, 62 of which are redder than the \citet{Meibom2011} sample. We analyze the pre-search data conditioning simple aperture photometry \citep[PDCSAP;][]{pdcsap1,pdcsap2} light curves from Quarters 2-16. Quarters 1 and 17 were truncated and only lasted for 33.47 d and 31.75 d, respectively, whereas the intervening Quarters covered $\approx$90 d, except for Quarter 8, which lasted for $\approx$67 d.
Most stars have more than one quarter of data.

\begin{figure*}[h!]\begin{center}
\includegraphics[trim=0.5cm 0cm 0.5cm 0cm, clip=True,  width=6.5in]{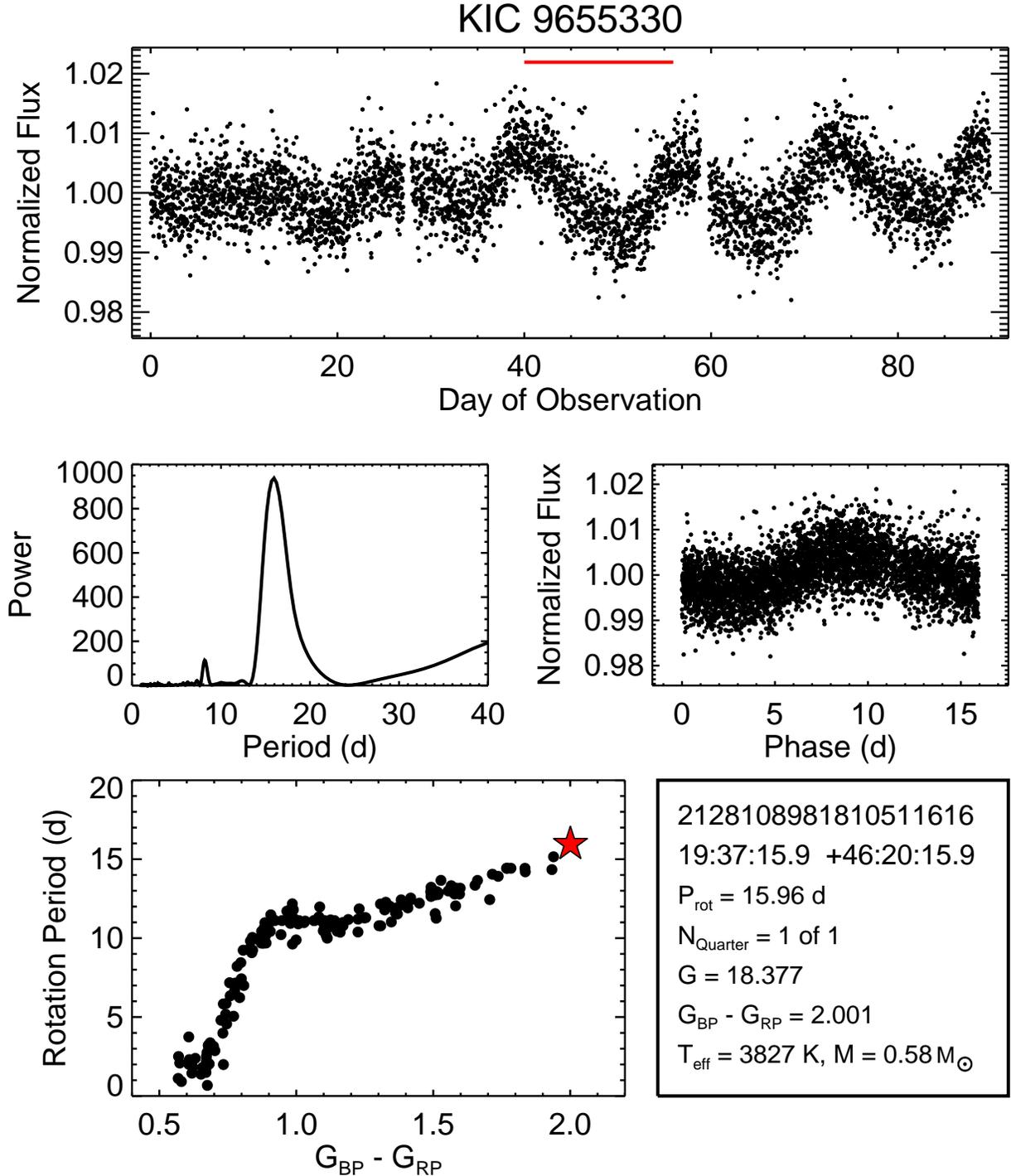} 
\caption{Analysis of the \textit{Kepler} light curve for KIC 9655330 (Gaia DR2 2128108981810511616), the 
faintest, reddest, and slowest star in our sample.
  \textit{Top---}The Quarter 6 \textit{Kepler} PDCSAP
  light curve for this target. 
    The length of the red line at the top left is the duration of one cycle (i.e., \prot).  
  This is the only available {\it Kepler} light curve for this star.
    \textit{Middle left---}The Lomb--Scargle periodogram for the light curve peaks at \prot\ = 15.96 d. 
  \textit{Middle right---}The phase-folded light curve shows a clean repeating pattern. 
  \textit{Bottom left---}The color--period diagram for NGC~6811
  with this star highlighted (red star).
  \textit{Bottom right---}Reference information for this star, 
  including Gaia DR2 Source ID, coordinates, 
  \prot\ (and the standard deviation of values when multiple quarters are available), 
  the number of quarters used to calculate \prot, 
  $G$ magnitude and \gbr\ color, and 
  estimates for \teff\ and mass.
  Versions of this figure panel for every target analyzed (171 images) are available as an electronic figure set in the online Journal.
       \label{f:ex}}
\end{center}\end{figure*}

We compute Lomb--Scargle periodograms \citep{Scargle1982, press1989} for all available quarters for each star (see Figure~\ref{f:ex} for an example of our analysis), and report the median and standard deviation values for the resulting \prot\ in Table \ref{table}. Occasionally, the peak power in the periodogram corresponds to a half-period harmonic (9\% of the total number of all \prot\ measurements). We automatically detect these cases by identifying outliers with values that when doubled were within 10\% of the 
median value from all quarters, then correct the measured \prot\ by doubling their values.

\begin{figure}[t]\begin{center}
\includegraphics[trim=1.2cm 0cm 0.4cm 1.5cm, clip=True,  width=2.75in]{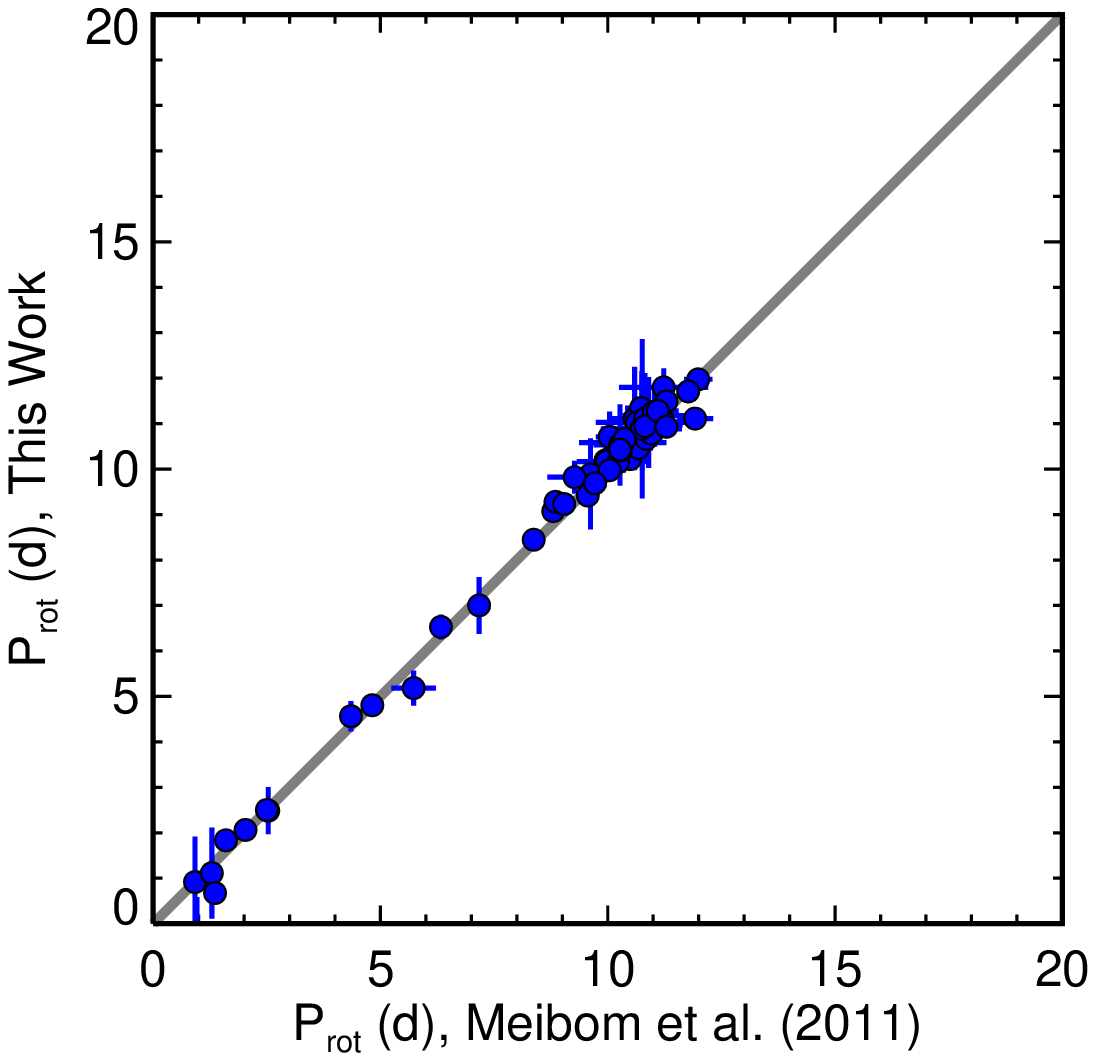} 
\includegraphics[trim=1.2cm 0cm 0.4cm 1.5cm, clip=True,  width=2.75in]{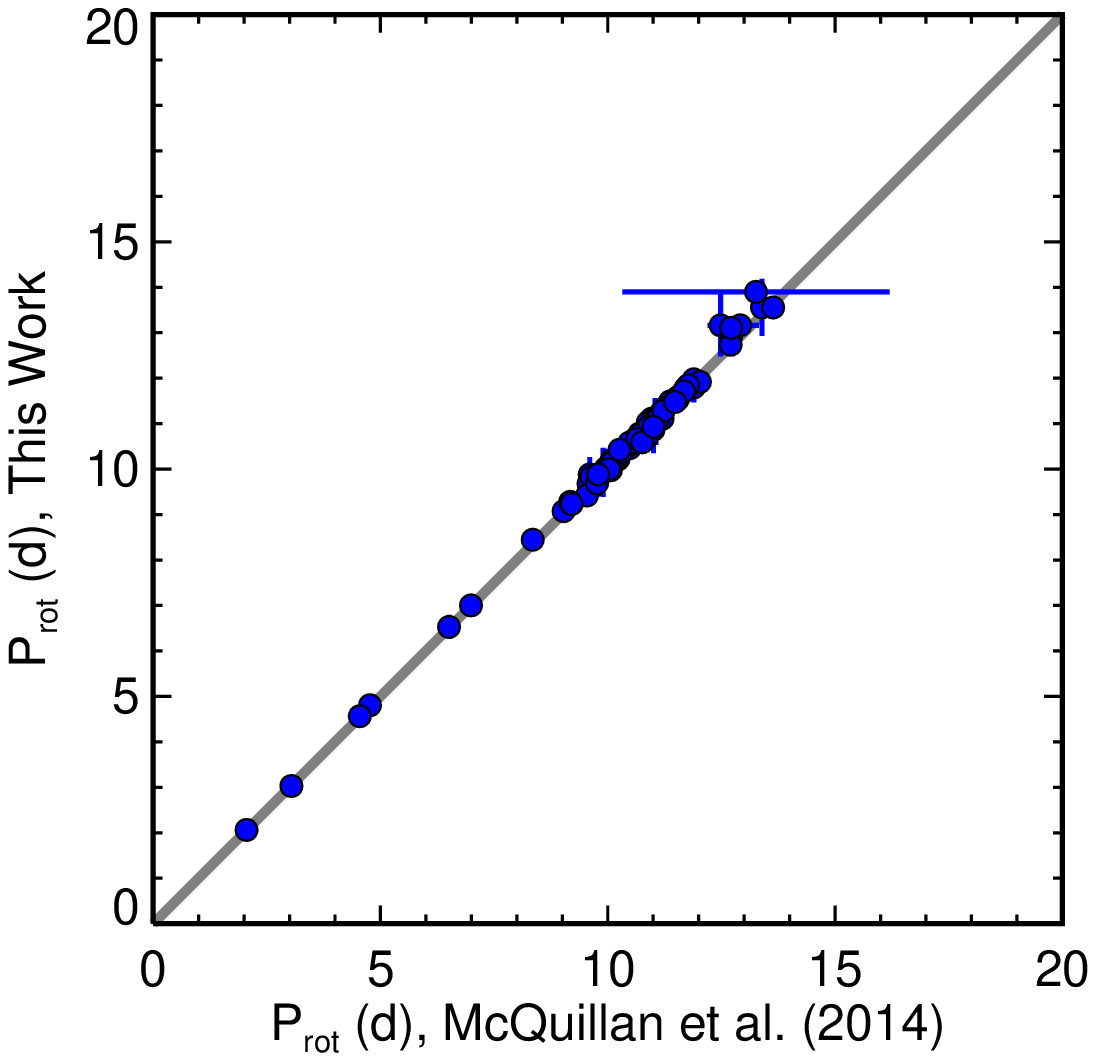} 
  \caption{Comparison between our measured periods and values from 
  (\textit{top}) \citet{Meibom2011} and 
  (\textit{bottom}) \citet{AmyKepler} for these same stars. Our measurements show excellent agreement with previous ones. 
    \label{f:litcompare}}
\end{center}\end{figure}

\citet{AmyKepler} reported \prot\ for 87 of our stars, 57 of which were not identified as members by \citet{Meibom2011}. 
In Figure \ref{f:litcompare}, we compare the \prot\ values we find to those in both \citet{Meibom2011} and \citet{AmyKepler}, 
and find close agreement. 
For \citet{AmyKepler}, we calculate the differences from our periods 
and find no net offset and a standard deviation of only $\sigma = 0.15$~d.

Finally, there are 11 stars with multi-quarter data with large standard deviations;
$\sigma \prot > 1$~d. 
We discard these from our sample and attribute them to limitations in the data quality or perhaps blending of stars in the \textit{Kepler} photometric apertures.
Our measurements are displayed in a color--\prot\ diagram in Figure \ref{f:6811}. Table \ref{table} lists the 171 candidate single-star members that now have a measured \prot, and includes 10 stars that are period outliers 
\citep[they appear too fast or too slow relative to the cluster color--\prot\ sequence, which we expect to be fully and tightly converged since the distributions for Praesepe and the Hyades are, as demonstrated by][]{Douglas2019}. The remaining 161 rotators more than double the size of the \citet{Meibom2011} sample and extend it to $\approx$0.6~\msun.

\clearpage
\section{Discussion} \label{s:dis}
Below, we calculate the reddening and age for NGC 6811 using gyrochronology. 
Next, we show that the NGC~6811 K dwarfs have not spun down relative to their cousins in Praesepe, despite their large difference in age. Finally, we present evidence supporting the scenario whereby  stars stop spinning down, effectively stalling in their angular-momentum evolution, for an extended period of time.

\subsection{A gyrochronology reddening and age for NGC 6811 using F and G members}
\label{s:red}

\begin{figure*}\begin{center}
\includegraphics[trim=1.5cm 0cm 0.0cm 0cm, clip=True,  width=5.in]{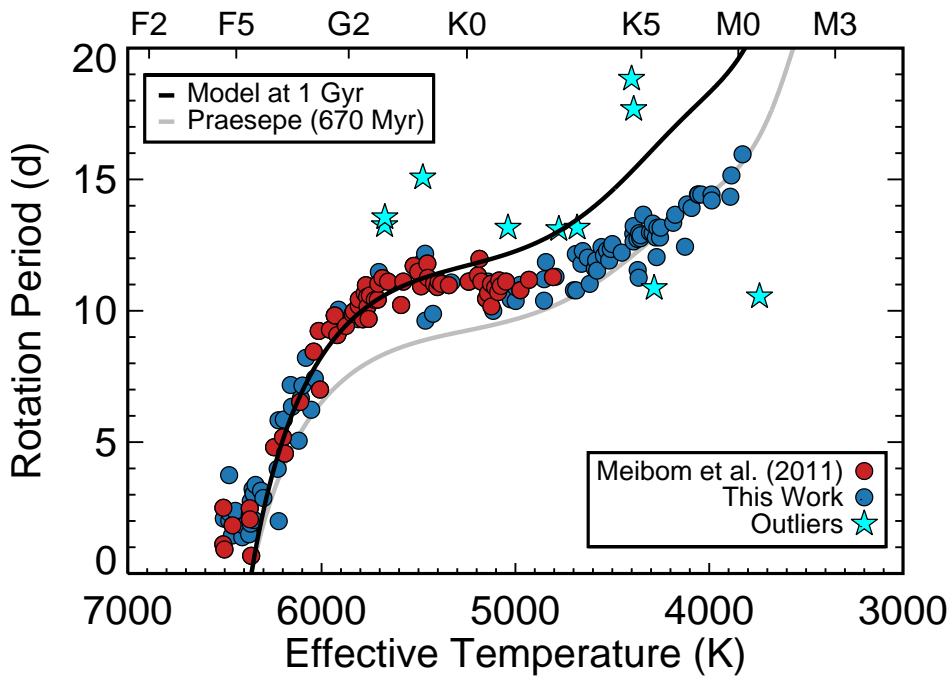} 
\caption{The effective temperature--period distribution for 171 members of NGC 6811. 
The 63 red-shaded points are from \citet{Meibom2011}, 
and the 10 cyan stars mark outliers with periods deviating from the main trend by more than 2 d. Our expanded sample more than doubles the number of known rotators and extends their mass range from 0.8 \msun\ to 0.6 \msun\ (blue points). We compare the NGC 6811 period sequence to a fit to Praesepe (lower line), which we also evolved forward to 1~Gyr with $n = 0.62$.  Stars with $\teff > 5400$~K
have all spun down relative to Praesepe following this model, but the sequence flattens out toward cooler temperatures, 
diverging from the NGC 6811-age model (upper line) and continuing toward the Praesepe sequence. 
 The stalling of spin-down 
  is now very clear: the lowest mass stars in NGC 6811 seem to be spinning at the same rates as their younger cousins in Praesepe.
\label{f:6811}}
\end{center}\end{figure*}

In addition to the \prot\ distribution for NGC~6811, Figures~\ref{f:mei} and \ref{f:6811} include information on the rotational properties of younger stars in Praesepe. In each figure, the lower gray line corresponds to the \citet{Douglas2019} fit to Praesepe's slowly rotating sequence. These authors constructed a gyrochronology model using these data as a first epoch (670 Myr, $A_V = 0.035$), and the Sun as a second epoch (4.567 Gyr, $\gbr_\odot = 0.82$), 
and derive a braking index $n = 0.62$.\footnote{According to 
models by \citet{vanSaders2013}, metallicity is not expected to significantly impact angular momentum evolution when period distributions 
are analyzed with respect to temperature (and therefore color) instead of mass.}
The second, higher gray line in each figure projects their fit from 670~Myr to 1 Gyr using this new braking index.

Inspecting the color--\prot\ sequence for NGC 6811 and the fit to Praesepe data shown in Figure~\ref{f:6811}, it is clear that the F and G stars in Praesepe (which sit on the lower gray line) have spun down by the age of NGC 6811. This means that we can use these stars to obtain an age for 
NGC 6811, as long as our gyrochronology model takes Praesepe as its first epoch.

\begin{figure*}
\begin{center}
\includegraphics[trim=0.4cm 0cm 0.4cm 0cm, clip=True,  width=2.3in]{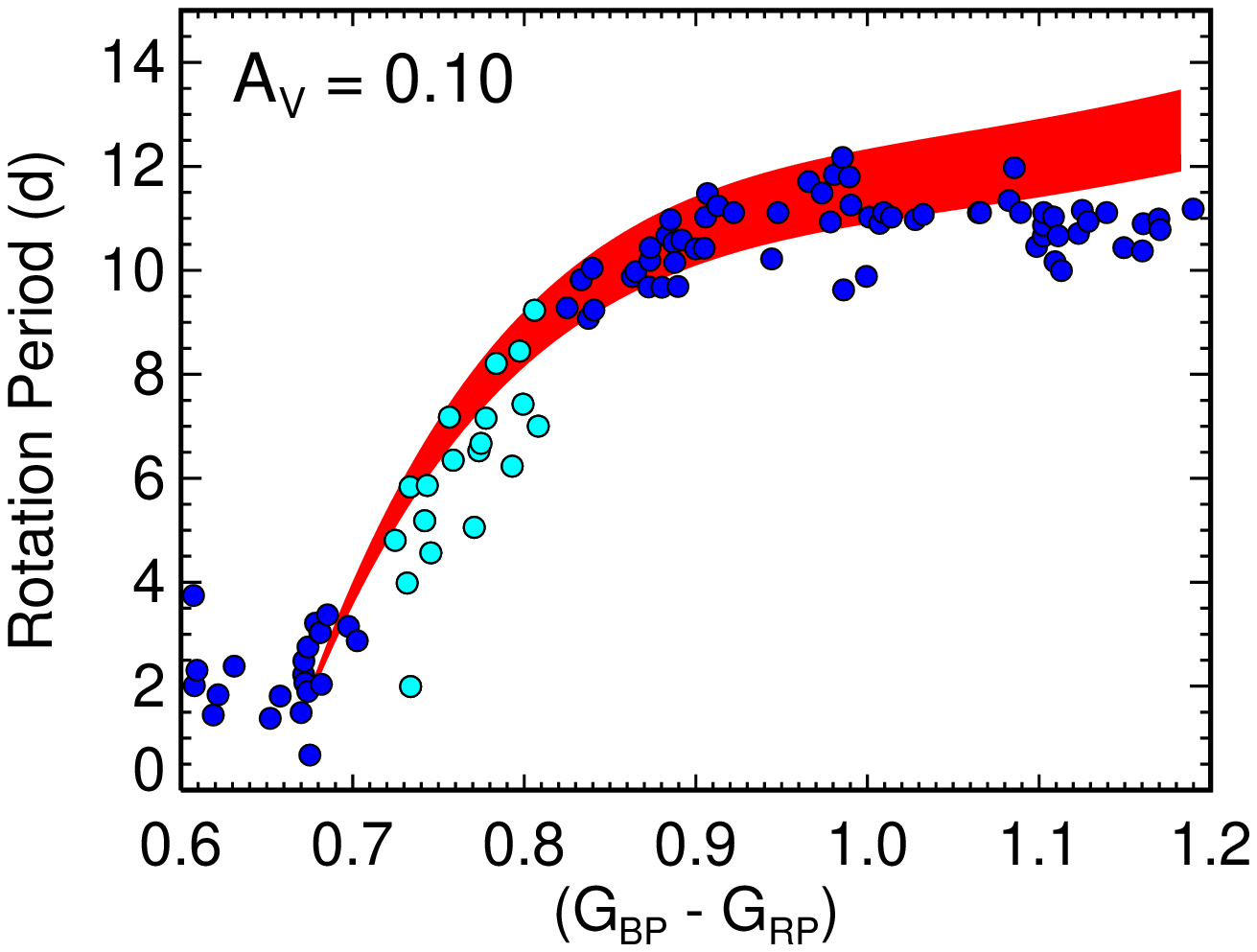} 
\includegraphics[trim=0.4cm 0cm 0.4cm 0cm, clip=True,  width=2.3in]{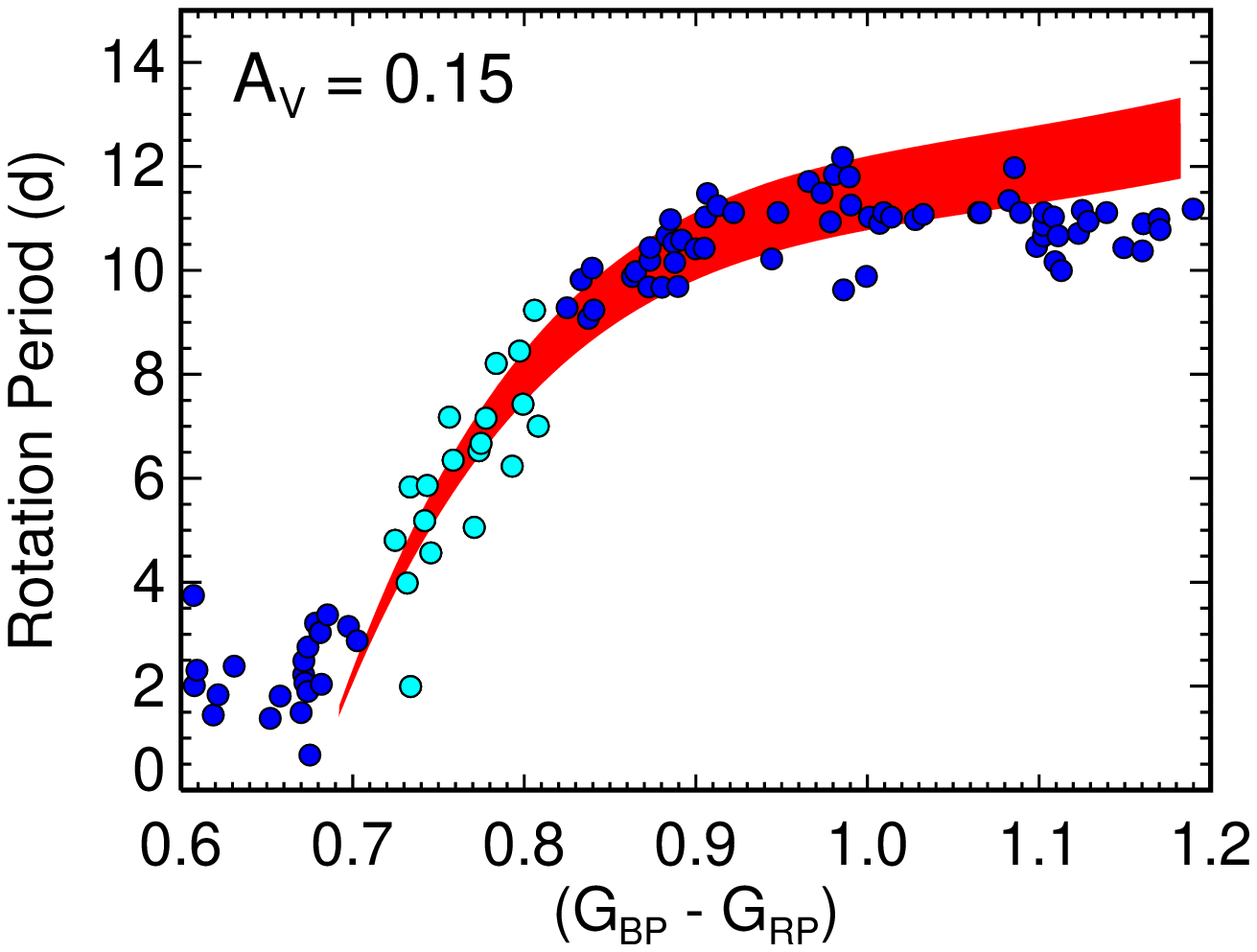} 
\includegraphics[trim=0.4cm 0cm 0.4cm 0cm, clip=True,  width=2.3in]{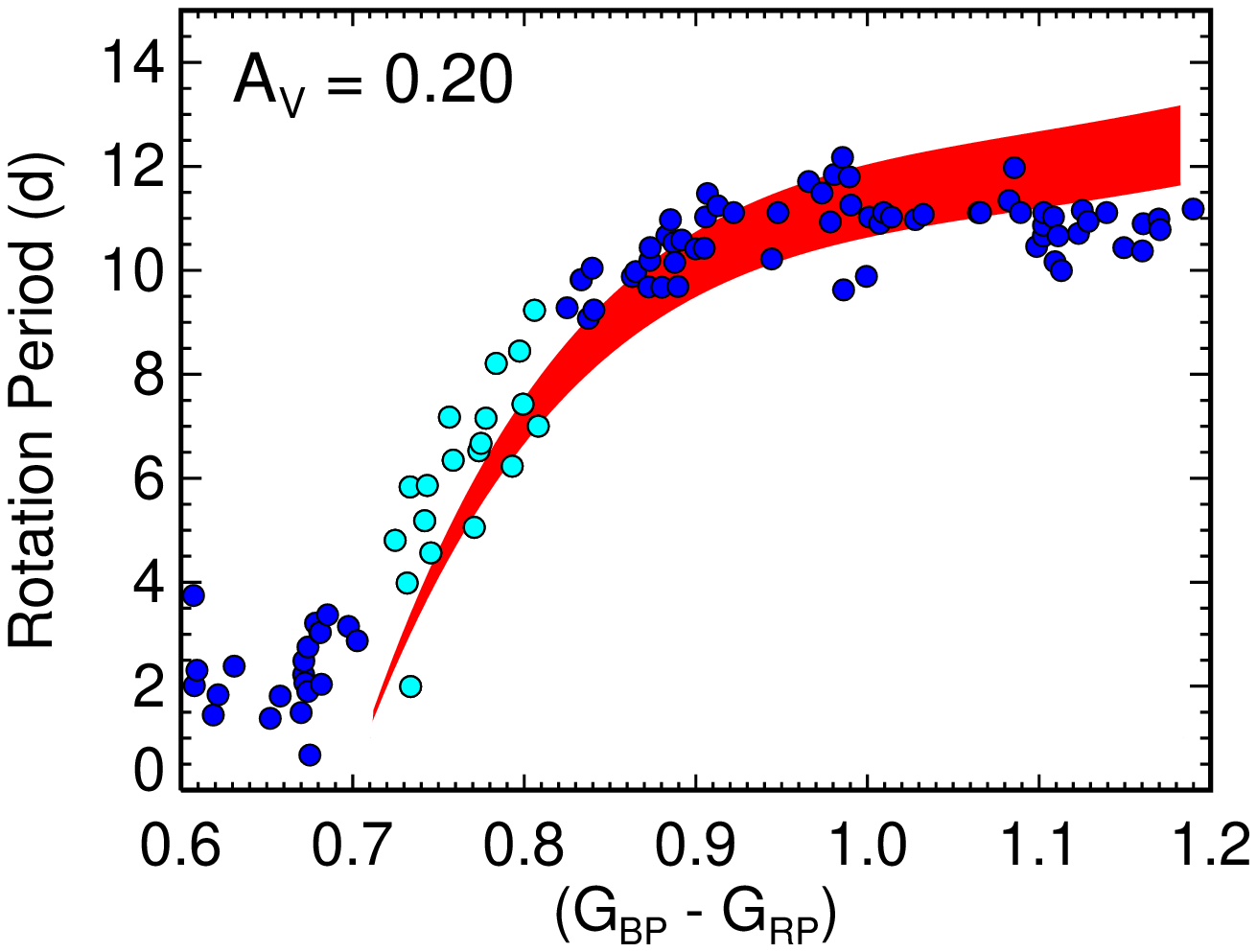} 
\caption{Interstellar reddening can be precisely constrained from the color--period diagram of a cluster with an approximately known age. The NGC 6811 color--period diagram is shown in all three panels, and the F dwarfs are highlighted (cyan points). Gyrochronology models built with Praesepe and the Sun ($n = 0.61$) with ages ranging from 0.9 to 1.1 Gyr are shaded in red, with extinction values applied of $A_V = 0.10$ (\textit{left}), 0.15 (\textit{middle}), and 0.20 (\textit{right}). The location of the F dwarf sequence is more sensitive to reddening than to age. Fitting our Praesepe-derived model to NGC 6811, we find $A_V = 0.15 \pm 0.02$, where the uncertainty comes from varying Praesepe's age between 650 and 700 Myr and NGC 6811's age between 0.9 and 1.1 Gyr.
    \label{f:red}}
\end{center}\end{figure*}

Reddening is an important ingredient in a gyrochronology age calculation. The rapid rise in F dwarf \prot\ over the narrow range in \teff\ seen in Figure~\ref{f:6811} (which corresponds to $0.65 < \gbr_0 < 0.75$, where $\gbr_0$ is the unreddened color; see Figure~\ref{f:red}) provides a tight constraint on reddening that is relatively insensitive to age. We fit for reddening by comparing the color--\prot\ sequences for Praesepe and NGC 6811. To account for Praesepe's uncertain age, we re-model the un-reddened Praesepe sequence with a sixth order polynomial, $A_V = 0.035$ and ages~$t = 650$ and 700 Myr, and re-calculate the braking index relative to the Sun, finding $n = 0.61$ and 0.63 for the two ages.

We project our Praesepe model forward to 1 Gyr, the isochrone age of NGC 6811, 
and minimize $\chi^2$ between our measured \prot\ and the model predictions for the 16 members with $0.65 < \gbr_0 < 0.75$. We repeat this exercise for ages of 0.9 and 1.1 Gyr to constrain the age sensitivity on reddening, and find 
$A_V = 0.156_{-0.020}^{+0.016}$ and 
$A_V = 0.145_{-0.023}^{+0.018}$, for 
$E(B-V) = 0.050_{-0.006}^{+0.005}$ and 
$E(B-V) = 0.047_{-0.007}^{+0.006}$, 
for the 650 and 700 Myr cases respectively. The exact ages for Praesepe and NGC 6811 only have a small impact on the resulting reddening, as we illustrate in Figure \ref{f:red}. 

Indeed, even adopting an 800~Myr age for Praesepe 
\citep{Brandt2015b}, 
the braking index increases to $n = 0.68$, the 
gyrochronology age for NGC~6811 increases to 1.2~Gyr, 
but we still find the same reddening value.
However, this age is inconsistent with the cluster CMD. 
A 1.2~Gyr PARSEC isochrone with solar metallicity can 
only be approximately fit to the cluster turnoff by 
reducing the reddening to zero.
This is inconsistent with expectations from the 3D dust map 
\citep{Green2018}, which has $A_V = 0.25 \pm 0.06$ at the location 
and distance of NGC~6811.
That value is larger than our result.
However, if $A_V = 0.25$, 
we would then infer an age of 1.26~Gyr with our gyrochronology model,
but then a CMD isochrone analysis would demand 
a significantly younger age of $\approx0.95$~Gyr.    
The only reddening and age combination 
that gives a precise and consistent match between isochrone and 
gyrochronology models with the data is 
our solution of $A_V = 0.15$ and 1~Gyr.
We use $A_V = 0.15$ for the remainder of this work.

To calculate the cluster age, we consider only G dwarfs with $0.75 < \gbr_0 < 0.90$ (29 stars within 
$5500 < \teff < 6000$~K)
by applying $A_V = 0.15$. 
While we do demonstrate in this paper that gyrochronology fails to describe K dwarf spin-down, 
\citet{Meibom2015} showed that gyrochronology does accurately 
represent G dwarfs at 2.5~Gyr and 
\citet{vanSaders2016} showed that it remains valid up to the 
age of the Sun.
We chose our color cutoff by inspecting where the model diverges 
from the data in Figure~\ref{f:6811}, corresponding to 
$\teff \approx 5400$~K.
We find a gyrochronological age of $t_{\rm gyro} = 1.04 \pm 0.12$~Gyr (median and standard deviation) if Praesepe is 670~Myr old (the age derived from an analysis of literature ages by \citet{Douglas2019}. Trimming the outliers shown in Figure~\ref{f:6811} (two of the cyan stars) 
reduces the standard deviation to 75~Myr or $\approx$9\%, which is incredibly precise. The standard deviation of the mean is only $\approx$1.5\%.
This gyrochronology age is consistent with the literature isochrone ages \citep[][]{Sandquist2016, Janes6811}.

As a final test, we solve for reddening and age simultaneously. We minimize $\chi^2$ by brute force with both properties as free parameters and find an age of 
1.00~Gyr and $A_V = 0.15$, 
consistent with our previous results.

\subsection{Further evidence for a temporary epoch of stalled spin-down for K stars}
\label{s:stall}

The \citet{Meibom2011} color--\prot\ sequence for NGC 6811 (Figure~\ref{f:mei}) seems oddly flat compared to that for other clusters and to model expectations: all of the stars less massive than $M_\star \lesssim 1$ \msun\ appear to have the same \prot. \citet{Meibom2011} noted that extrapolating a younger cluster forward in age with the Skumanich Law predicts a different, more sloped shape to this sequence, with lower-mass stars rotating 
more slowly than their more massive neighbors in color--\prot\ space.

\begin{figure*}\begin{center}
\includegraphics[trim=0.6cm 0cm 0.0cm 0cm, clip=True,  width=3.5in]{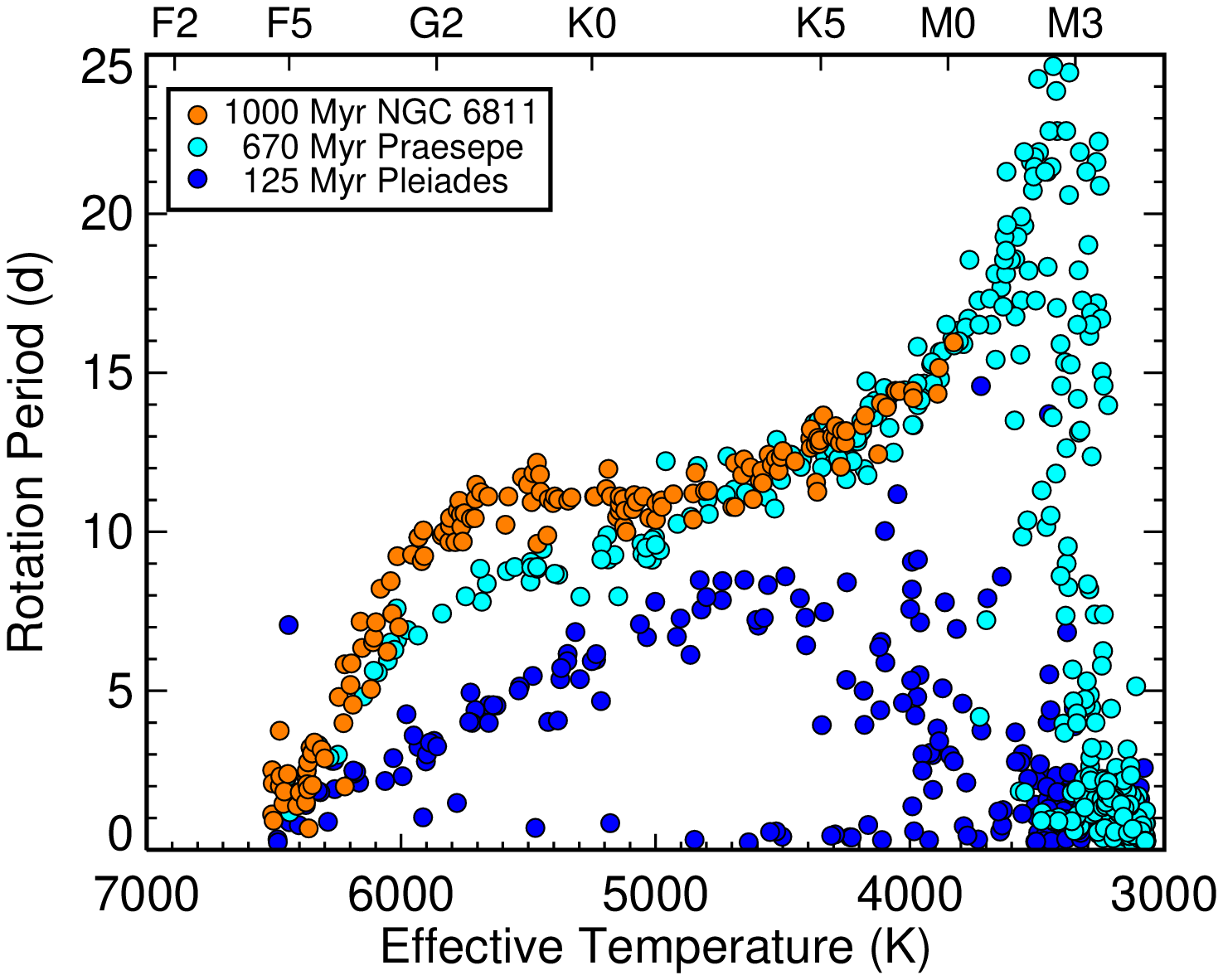} 
\includegraphics[trim=0.6cm 0cm 0.0cm 0cm, clip=True,  width=3.5in]{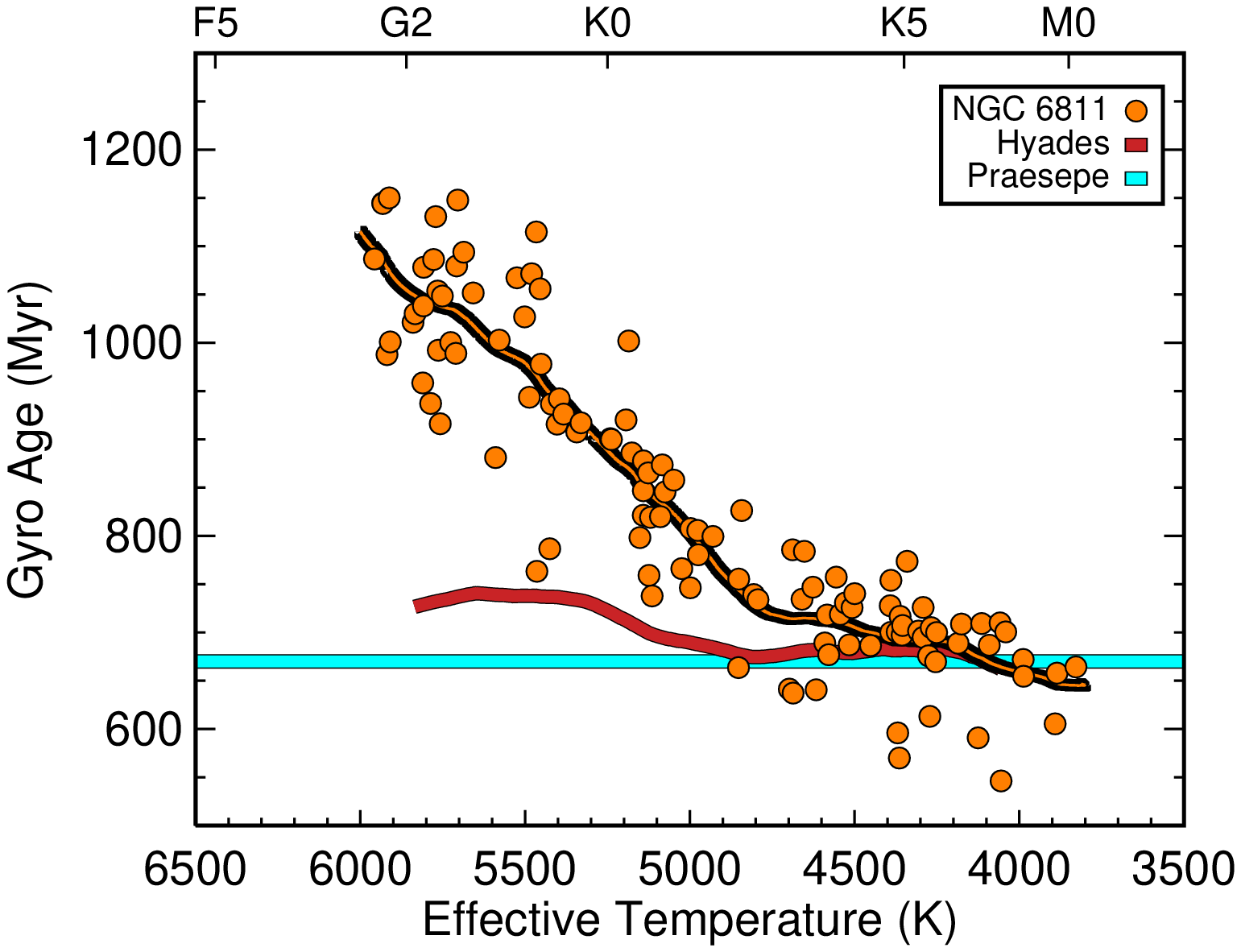} 
\caption{\textit{Left---}The effective temperature--period distributions for the Pleiades \citep[blue points, 125 Myr;][]{Rebull2016}, Praesepe \citep[cyan points, 670 Myr;][]{douglas2017, Douglas2019}, and NGC 6811 (orange points). While stars in the Pleiades hotter than $\approx$4000~K have clearly spun down by the age of Praesepe, those cooler than 5000 K appear to stall between the age of Praesepe and that of NGC 6811.
\textit{Right---}The differential gyrochronology ages for NGC 6811 rotators (orange points) compared to our fiducial Praesepe model (cyan horizontal line; 670~Myr). We also include the LOWESS regression for the age difference (black--orange line), showing that the difference in gyrochronology ages is strongest for G stars, and decreases between $5250 > \teff > 4700$~K until the cooler stars appear coeval with Praesepe.
The same analysis for the Hyades relative to Praesepe 
\citep{Douglas2019}
is shown as the red curve.
  \label{f:compare}}
\end{center}\end{figure*}

Our work confirms NGC 6811's departure from expectations,  illustrated in Figure~\ref{f:6811}. The rapid rise in \prot\ from the late-F dwarfs to the early G dwarfs is well represented by the 1 Gyr Praesepe projection. By contrast, the NGC 6811 sequence appears to diverge at redder colors and approach the Praesepe sequence. The natural conclusion is that K dwarfs have a lower braking index. 

This challenge to the picture of a color-independent $n$ is not  unexpected: \citet{Angus2015} found that NGC 6811 and Praesepe had color dependencies that were different from each other and from the 
Hyades and Coma Ber clusters, leading these authors to discard the NGC 6811 and Praesepe samples from their calibration sample. \citet{Angus2015} speculated that these fitting problems could indicate limitations in the gyrochronology formula, and listed metallicity as a possible factor.

\citet{Agueros2018} suggested an alternative scenario based on the  results for NGC 752, whose age of 1.4~Gyr makes it significantly older than NGC 6811 and Praesepe. \citet{Agueros2018} found that five mid-K dwarfs in the cluster had barely spun down compared to their cousins in Praesepe, while the three late-K/early M dwarfs had not slowed at all.\footnote{\citet{Agueros2018} reported rotation periods for 12 stars in NGC 752. Four of these have \textit{Gaia} DR2 astrometry and/or photometry inconsistent with single-star membership, and so we discarded them for the present discussion.} \citet{Agueros2018} hypothesized that stars enter a temporary phase of reduced net braking efficiency, where the stars stop spinning down. Based on where the NGC 6811 and NGC 752 color--\prot\ sequences merged together with Praesepe's, \citet{Agueros2018} inferred that the duration of the stalling epoch must depend on mass, and last longer for lower-mass stars.

This work led us to predict that if the color--\prot\ sequence for NGC 6811 were extended to lower masses, it would eventually merge  with Praesepe's. Indeed, as shown in Figure~\ref{f:6811}, 
NGC 6811's deviation away from the 1 Gyr projection continues until $\teff \approx 4295$~K (K6V), where it clearly overlaps with Praesepe. The late-K dwarfs of NGC 6811 have not spun down at all over the $\approx$350 Myr separating the two clusters.

This conclusion is even more striking when data for a younger cluster are included. The left panel of Figure~\ref{f:compare} shows \prot\ for the Pleiades \citep[125 Myr;][]{Rebull2016}, 
Praesepe \citep[][]{douglas2017,Douglas2019}, 
and NGC 6811.
The Pleiades K dwarfs with $4500 < \teff < 5000$~K have slowed down appreciably by 670~Myr (stars cooler than 4500~K yet converged on the slow sequence). However, the NGC 6811 stars in this same temperature range have hardly spun down relative to Praesepe, and the late-K/early M dwarfs have not spun down at all, despite their large difference in age. 

The right panel of Figure~\ref{f:compare} 
shows gyrochronology ages for 
individual stars in NGC~6811.
While the G dwarfs show that the cluster is clearly older than Praesepe (represented by the cyan horizontal line at 670~Myr), this  difference in age between NGC 6811 and Praesepe appears to vanish at $\teff \lesssim 4700$~K.

\subsection{Validating the stalled braking scenario with \textit{Kepler}}
\label{s:kep}
While 
in models $n$ is generally taken to be independent of color and constant in time  \citep[e.g.,][]{barnes2007}, data like those presented in Figure~\ref{f:6811} have been used to suggest that K dwarfs spin down more gradually than their more massive siblings. For example, comparing the \prot\ sequences of M34 and NGC 6811 to that for the Hyades led \citet{MeibomM34, Meibom2011} to suggest that $n$ depends on color and is smaller for redder/cooler stars. 

\begin{figure}\begin{center}
\includegraphics[trim=0.75cm 0cm 0.0cm 0cm, clip=True,  width=3.4in]{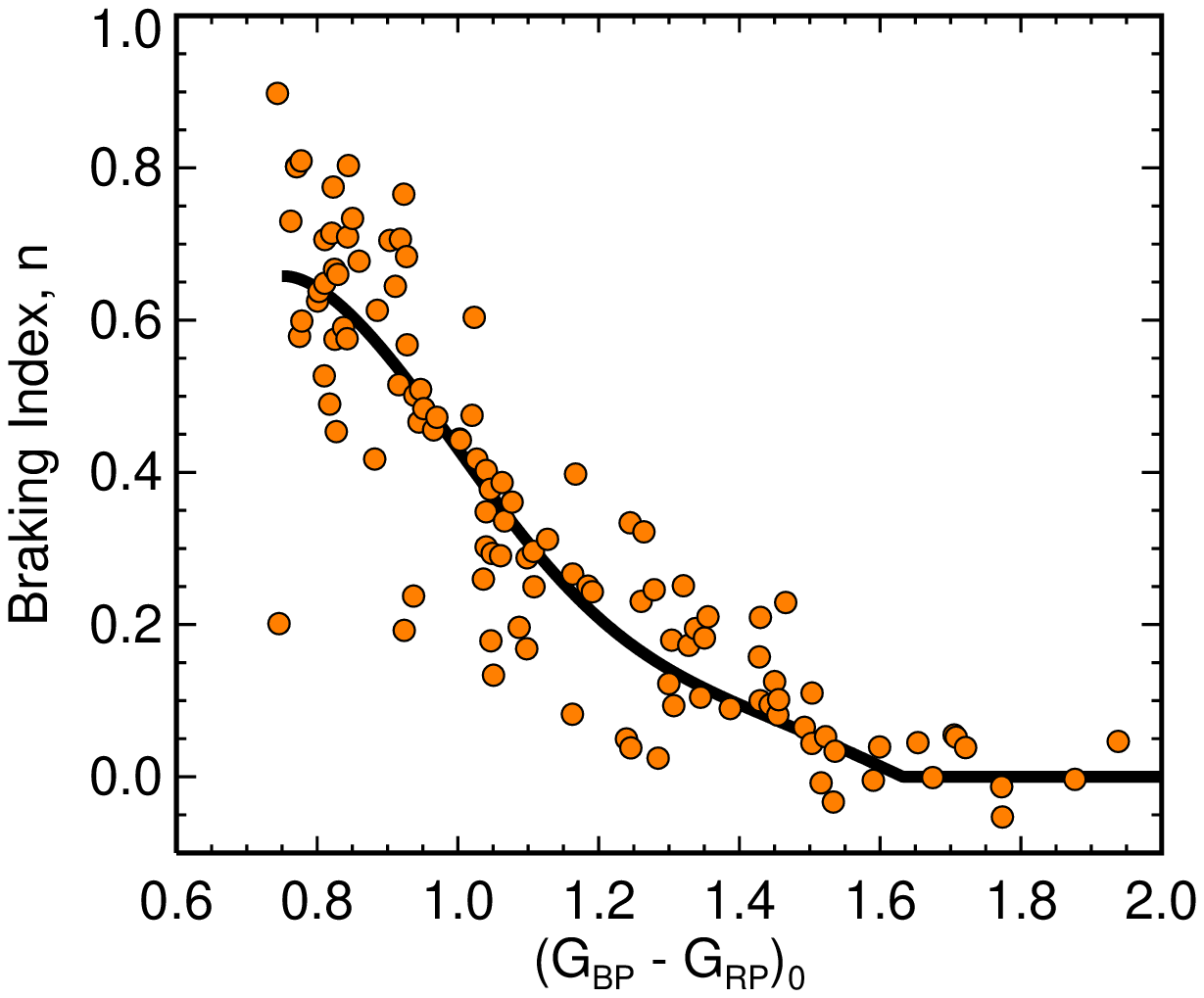} 
\includegraphics[trim=0.75cm 0cm 0.0cm 0cm, clip=True,  width=3.4in]{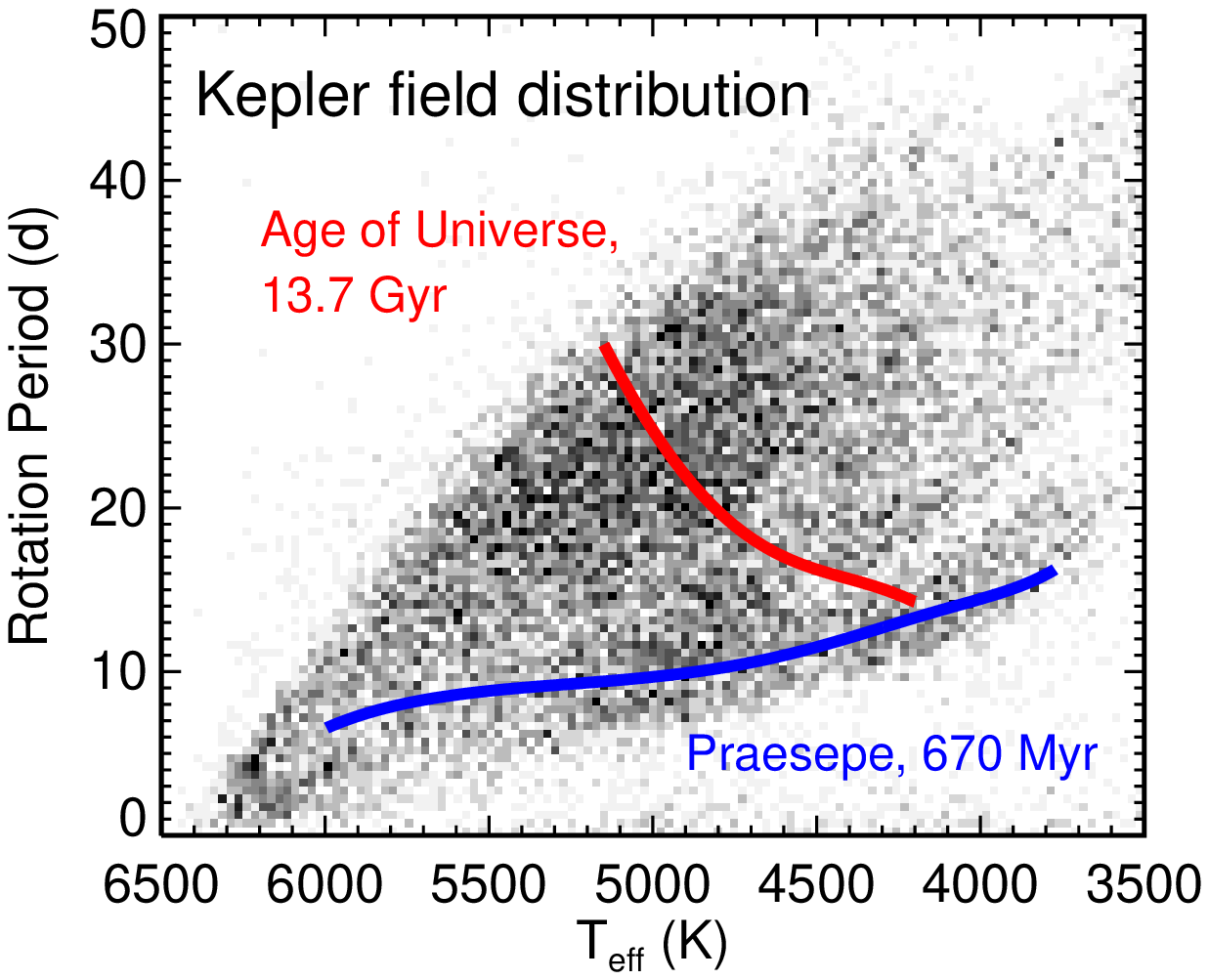} 
\caption{\textit{Top---}The braking index $n$ is calculated for NGC 6811 stars relative to Praesepe as a function of color. This plot could be interpreted as showing that spin-down depends on color. 
\textit{Bottom---}The \textit{Kepler} \prot\ distribution \citep{AmyKepler} is plotted along with the Praesepe fit (670 Myr, blue line), and a 13.7 Gyr model created by projecting the Praesepe fit forward with the $n$ from the top panel (red line). Clearly, a color-dependent braking law tuned with Praesepe and NGC 6811 cannot explain the observed \textit{Kepler} \prot\ distribution. The Universe is not old enough for those K dwarfs to have spun down to $\prot \approx$ 30-40~d. The model is only plotted for K dwarfs, 
since more massive stars have lifetimes less than the age of the Universe.
    \label{f:ndex}}
\end{center}\end{figure}

We test this hypothesis 
by calculating a color-dependent braking index using our data for NGC 6811 and our empirical fit to Praesepe: if $\prot \propto t^n$, then $n = \log(P_{\rm rot, 2} / P_{\rm rot, 1}) / \log(t_2 / t_1)$.
We adopt ages for Praesepe and NGC 6811 of 670 Myr and 1 Gyr, and $A_V = 0.035$ and 0.15, respectively. The resulting dependence of $n$ on DR2 color is shown in Figure \ref{f:ndex}. By eye, it seems that $n$ is $\approx$constant for $\gbr_0 < 0.9$, after which it drops down toward 0.

It appears that the NGC 6811 data support a reduced braking efficiency for K dwarfs. However, this explanation cannot be correct, as we show by testing this model with the  \textit{Kepler} \prot\ distribution for field stars \citep{AmyKepler}. We propagate the Praesepe color--\prot\ sequence forward in time, assuming a color-dependent $n$, and compare the results to the observed distribution of \prot. The bottom panel of Figure \ref{f:ndex} shows the {\it Kepler} \prot\ as a function of \teff, along 
with the fit for Praesepe and the prediction for the approximate age of the Universe, 13.7~Gyr. According to this model, the Universe is not old enough for the K dwarfs in the \textit{Kepler} field to have spun down to their observed $\prot \approx$ 30-40~d. 

We therefore interpret the data in the left panel of Figure~\ref{f:compare} differently. Following the \citet{Agueros2018} hypothesis, we argue that spin-down stalls after stars converge on the slow sequence for some amount of time, 
after which stars resume braking as expected with a common $n$. The shape of the curve in the right panel of Figure~\ref{f:compare} reveals how the duration of stalling increases toward lower masses and cooler $\teff$.

\section{Conclusions}\label{s:concl}

Prior to this work, the sample of rotators in the 1-Gyr-old cluster NGC~6811 was limited 
to 71 RV-confirmed members with masses $\gtrsim$0.8~\msun. 
Fortunately, many more candidate members were targeted for observation for the \textit{Kepler} Cluster Study \citep{Meibom2011}.

We used data from \textit{Gaia} DR2 to identify hundreds more likely single members 
with \textit{Kepler} light curves, 
and we measured rotation periods for 171 of them.
This more than doubles the size of the rotator sample for this cluster, 
and importantly extends it down to $M_\star \approx$~0.6~\msun, 
covering the full K dwarf range.

Focusing on the G dwarfs, for which we expect the Skumanich Law to be valid at least up to the age of the Sun \citep{vanSaders2016}, we find an extremely precise gyrochronological age of 
$t_{\rm gyro} = 1.04 \pm 0.07$~Gyr
(median and standard deviation of 27 G-type stars), 
relative to the 670-Myr-old benchmark Praesepe 
(i.e., $\approx$$1.5\times$ older).
However, this difference in age appears to vanish for stars 
cooler than $\teff < 4800$~K.

This could be interpreted as evidence 
for a mass-dependent braking index $n$, 
resulting in K dwarfs spinning down much more slowly than F and G dwarfs.
However, this scenario cannot reproduce the distribution of \prot~$\approx$~30-40 d observed for these stars in the \textit{Kepler} field by \citet{AmyKepler}---the Universe is not old enough for K dwarfs to spin down to these $\prot$ under this model.

Instead, we argue that K dwarfs stop spinning down after 
converging on the slow sequence at some time prior to the age of Praesepe. Braking stalls for an extended period of time, 
the duration of which increases toward lower masses/cooler temperatures. 
At some age older than NGC~6811 (1~Gyr) or NGC~752 (1.4~Gyr),
the braking efficiency increases to a Skumanich-like value, 
and the stars resume spinning down.

What might be causing this epoch of stalled braking? The cluster data demonstrate that the net angular-momentum loss is low during this phase compared to the time before and following it. If this is the case, then either the braking torque is temporarily reduced, and/or the photosphere is gaining angular momentum from the interior to offset that lost via magnetic braking 
\citep[e.g., 
][]{Hartman2010, Denissenkov2010, Bouvier2008, Gallet2013, Gallet2015, Lanzafame2015}.

\citet{Denissenkov2010} discuss a core--envelope decoupling model, 
and quote a timescale of $\tau_{\rm c-e} = 55 \pm 25$~Myr and $175 \pm 25$~Myr for 1.0~\msun\ and 0.8~\msun\ stars
($\teff \approx 4850$~K, K2).
This K dwarf timescale does not seem long enough to address the spin-down discrepancy 
we see between NGC 6811 and Praesepe. 
\citet{Gallet2015} quote values of $\tau_{\rm c-e}$ = 150--500~Myr for 0.5~\msun\ stars, 
but their slowest track is only $\approx70\%$ the period of Praesepe's converged slow sequence, so modeling challenges persist.
We are eager to see these theoretical models recalibrated with our new rotator sample for NGC 6811.

We propose that measurements of rotation in an even older cluster, 
for example, Ruprecht 147 \citep[2.5 Gyr;][]{Torres2018, Curtis2016PhD, Curtis2013},
will show that the K dwarfs resume spinning down more efficiently after the 
age of NGC 6811 or NGC 752, 
and we expect that the braking 
index that is found will be consistent with the 
distribution of \prot\ observed in the \textit{Kepler} field.

\acknowledgments

J.L.C. is supported by the National Science Foundation 
Astronomy and Astrophysics Postdoctoral Fellowship under award AST-1602662 and
the National Aeronautics and Space Administration under
grant NNX16AE64G issued through the \textit{K2} Guest Observer Program (GO 7035).
M.A.A. acknowledges support provided by the NSF through
grant AST-1255419.
S.T.D.~acknowledges support provided by the NSF through grant AST-1701468.
We thank Jennifer van Saders and Florian Gallet for sharing angular momentum evolution models 
\citep{vanSaders2013, Gallet2015},
and the anonymous referee for providing a helpful and thorough review of this manuscript.

This paper includes data collected by the \textit{Kepler} and 
\textit{K2} missions, 
which are funded by the NASA Science Mission directorate.
We obtained these data from the Mikulski Archive for Space Telescopes (MAST). 
STScI is operated by the Association of Universities for Research in Astronomy, Inc., 
under NASA contract NAS5-26555. 
Support for MAST for non-HST data is provided by the NASA Office of Space Science via 
grant NNX09AF08G and by other grants and contracts.

This work has made use of data from the European Space Agency (ESA)
mission {\it Gaia},\footnote{\url{https://www.cosmos.esa.int/gaia}} processed by
the {\it Gaia} Data Processing and Analysis Consortium (DPAC,\footnote{\url{https://www.cosmos.esa.int/web/gaia/dpac/consortium}}). Funding
for the DPAC has been provided by national institutions, in particular
the institutions participating in the {\it Gaia} Multilateral Agreement.

This research has also made use of NASA's Astrophysics Data System, 
and the VizieR \citep{vizier} and SIMBAD \citep{simbad} databases, 
operated at CDS, Strasbourg, France.

\facilities{Gaia, Kepler}


\software{The IDL Astronomy User's Library \citep{IDLastro}}


\bibliographystyle{aasjournal}

\clearpage
\startlongtable
\begin{deluxetable*}{rccccccccccc}
\tablecaption{Data for the NGC 6811 Benchmark Sample \label{table}}
\tablewidth{0pt}
\tablehead{
\colhead{\#} & \colhead{KIC ID} & \colhead{\textit{Gaia} DR2 Source ID} & \colhead{$G$} & \colhead{$(G_{BP} - G_{RP})$} & 
\colhead{\teff} & 
\colhead{Mass} & 
\colhead{SpT} & 
\colhead{\prot} & \colhead{$\sigma \prot$} & \colhead{$N_{\rm Q}$} & 
\colhead{Code} \\
\colhead{} & \colhead{} & \colhead{} & \colhead{(mag)} & \colhead{(mag)} & 
\colhead{(K)} & 
\colhead{(\msun)} & 
\colhead{} & 
\colhead{(d)} & \colhead{(d)} & \colhead{} & 
\colhead{}
}
\startdata
  1 &  9716563 & 2128134824634071424 & 13.313 & 0.569 & 6507 & 1.287 & F5 &  1.11 & 0.3380 & 15 & Y \\
  2 &  9716076 & 2128122008451548672 & 13.350 & 0.570 & 6506 & 1.287 & F5 &  2.50 & 0.0040 & 15 & Y \\
  3 &  9718403 & 2080469827339292672 & 13.444 & 0.574 & 6504 & 1.286 & F5 &  2.09 & 0.3015 & 15 & Y \\
  4 &  9715923 & 2128125135187725696 & 13.337 & 0.580 & 6501 & 1.284 & F5 &  0.92 & 0.0006 & 15 & Y \\
  5 &  9716858 & 2128133789541221888 & 13.581 & 0.607 & 6477 & 1.271 & F5 &  3.74 & 0.5527 & 15 & Y \\
  6 &  9777642 & 2128137229815871232 & 13.547 & 0.608 & 6476 & 1.270 & F5 &  2.01 & 0.2010 & 15 & Y \\
  7 &  9716139 & 2128145506212004608 & 13.435 & 0.609 & 6474 & 1.269 & F5 &  2.31 & 0.3108 & 14 & Y \\
  8 &  9898009 & 2128518829068874112 & 13.702 & 0.619 & 6462 & 1.263 & F5 &  1.44 & 0.1338 & 15 & Y \\
  9 &  9654924 & 2128121561774912128 & 13.635 & 0.622 & 6458 & 1.260 & F5 &  1.83 & 0.1558 & 15 & Y \\
 10 &  9655708 & 2128131216861127680 & 13.665 & 0.631 & 6445 & 1.253 & F5 &  2.38 & 0.3009 & 15 & Y \\
 11 &  9655437 & 2128132728689966720 & 13.812 & 0.652 & 6409 & 1.240 & F5 &  1.38 & 0.0767 & 15 & Y \\
 12 &  9655357 & 2128108951750969088 & 13.722 & 0.658 & 6398 & 1.236 & F5 &  1.81 & 0.1239 & 15 & Y \\
 13 &  9655145 & 2128121080738614016 & 13.905 & 0.670 & 6374 & 1.228 & F5 &  1.49 & 0.0025 & 15 & Y \\
 14 &  9594739 & 2128107714800080384 & 13.891 & 0.671 & 6370 & 1.227 & F5 &  2.22 & 0.1145 & 15 & Y \\
 15 &  9469799 & 2128018963596952832 & 13.896 & 0.672 & 6370 & 1.227 & F5 &  2.49 & 0.1439 & 15 & Y \\
 16 &  9655730 & 2128131216861484928 & 13.895 & 0.672 & 6369 & 1.226 & F5 &  2.06 & 0.0787 & 15 & Y \\
 17 &  9594038 & 2128118057081550464 & 13.944 & 0.674 & 6365 & 1.225 & F5 &  1.90 & 0.0528 & 15 & Y \\
 18 &  9716253 & 2128144857677611264 & 13.931 & 0.674 & 6365 & 1.225 & F5 &  2.75 & 0.2470 & 15 & Y \\
 19 &  9716817 & 2128133450244545024 & 13.801 & 0.675 & 6363 & 1.224 & F5 &  0.67 & 0.3636 & 15 & Y \\
 20 &  9777089 & 2128141524783121792 & 14.000 & 0.678 & 6356 & 1.222 & F5 &  3.21 & 0.1883 & 15 & Y \\
 21 &  9655727 & 2128132075854976896 & 13.975 & 0.681 & 6350 & 1.220 & F5 &  3.03 & 0.1512 & 15 & Y \\
 22 &  9591888 & 2128077168993658624 & 14.012 & 0.681 & 6350 & 1.220 & F5 &  3.03 & 0.1512 & 15 & Y \\
 23 &  9715848 & 2128124585431880704 & 13.959 & 0.682 & 6348 & 1.220 & F5 &  2.04 & 0.0204 &  9 & Y \\
 24 &  9655716 & 2128132075854974336 & 14.072 & 0.685 & 6340 & 1.217 & F5 &  3.37 & 0.2665 & 12 & Y \\
 25 &  9655385 & 2128109123549687680 & 14.035 & 0.698 & 6312 & 1.208 & F8 &  3.15 & 0.1494 & 15 & Y \\
 26 &  9595079 & 2128127265491233280 & 14.132 & 0.703 & 6300 & 1.203 & F8 &  2.87 & 0.0948 &  6 & Y \\
 27 &  9656480 & 2080097058538601088 & 14.318 & 0.725 & 6244 & 1.185 & F8 &  4.81 & 0.1290 & 15 & Y \\
 28 &  9533215 & 2128103179314615296 & 14.154 & 0.732 & 6226 & 1.179 & F8 &  3.98 & 0.2602 &  9 & Y \\
 29 &  9655911 & 2128128742959981312 & 14.228 & 0.734 & 6222 & 1.177 & F8 &  5.83 & 0.5253 & 15 & Y \\
 30 &  9716376 & 2128144960756839424 & 14.207 & 0.734 & 6221 & 1.177 & F8 &  1.99 & 0.0587 & 15 & Y \\
 31 &  9716401 & 2128145098195807872 & 14.205 & 0.742 & 6199 & 1.170 & F8 &  5.18 & 0.4928 & 15 & Y \\
 32 &  9655282 & 2128108539434091392 & 14.370 & 0.743 & 6195 & 1.168 & F8 &  5.86 & 0.2411 &  6 & Y \\
 33 &  9775854 & 2128126165979845376 & 14.278 & 0.746 & 6189 & 1.166 & F8 &  4.56 & 0.1771 & 10 & Y \\
 34 &  9715126 & 2128165817117806464 & 14.297 & 0.756 & 6159 & 1.156 & F8 &  7.18 & 0.6759 &  4 & Y \\
 35 &  9656016 & 2128130426587581440 & 14.459 & 0.759 & 6153 & 1.153 & F8 &  6.35 & 0.3323 & 13 & Y \\
 36 &  9593926 & 2128119190952918016 & 14.324 & 0.771 & 6118 & 1.141 & F8 &  5.05 & 0.2858 &  9 & Y \\
 37 &  9471344 & 2128098849987500160 & 14.515 & 0.774 & 6110 & 1.138 & G0 &  6.53 & 0.3318 & 15 & Y \\
 38 &  9716435 & 2128144926397113728 & 14.525 & 0.775 & 6106 & 1.137 & G0 &  6.67 & 0.2010 &  9 & Y \\
 39 &  9655268 & 2128120908939932032 & 14.520 & 0.778 & 6098 & 1.134 & G0 &  7.15 & 0.4493 & 15 & Y \\
 40 &  9776415 & 2128145957189232256 & 14.562 & 0.784 & 6080 & 1.128 & G0 &  8.21 & 0.6122 & 15 & Y \\
 41 &  9532828 & 2128102526479557248 & 14.528 & 0.793 & 6053 & 1.118 & G0 &  6.23 & 0.3795 & 15 & Y \\
 42 &  9777258 & 2128140425271501824 & 14.692 & 0.797 & 6040 & 1.114 & G0 &  8.45 & 0.1832 & 10 & Y \\
 43 &  9776409 & 2128151008070805888 & 14.709 & 0.799 & 6034 & 1.112 & G0 &  7.42 & 0.3478 &  5 & Y \\
 44 &  9778187 & 2080475084380577152 & 14.795 & 0.806 & 6014 & 1.105 & G0 &  9.23 & 0.4286 & 15 & Y \\
 45 &  9410241 & 2128096616604445312 & 14.679 & 0.808 & 6008 & 1.104 & G0 &  7.00 & 0.1357 & 15 & Y \\
 46 &  9776475 & 2128146472585332096 & 14.857 & 0.825 & 5957 & 1.089 & G0 &  9.28 & 0.6038 & 15 & Y \\
 47 &  9534045 & 2080086338300220032 & 14.904 & 0.833 & 5932 & 1.081 & G2 &  9.82 & 0.5624 & 15 & Y \\
 48 &  9656165 & 2128130186069010944 & 14.892 & 0.833 & 5931 & 1.081 & G2 &  9.82 & 0.5394 & 15 & Y \\
 49 &  9655917 & 2128131869696568960 & 14.899 & 0.837 & 5919 & 1.077 & G2 &  9.08 & 0.2432 & 15 & Y \\
 50 &  9411496 & 2080057785357418240 & 14.949 & 0.840 & 5912 & 1.075 & G2 & 10.04 & 0.4641 & 14 & Y \\
 51 &  9530806 & 2128068853937057408 & 15.030 & 0.841 & 5909 & 1.074 & G2 &  9.24 & $\cdots$ &  1 & Y \\
 52 &  9594287 & 2128106477849445248 & 14.904 & 0.863 & 5840 & 1.056 & G2 &  9.88 & 0.4857 & 15 & Y \\
 53 &  9716008 & 2128145785390512896 & 15.004 & 0.865 & 5833 & 1.054 & G2 &  9.98 & 0.4072 & 15 & Y \\
 54 &  9655424 & 2128108676873069952 & 14.915 & 0.873 & 5810 & 1.049 & G5 &  9.68 & 0.3383 & 15 & Y \\
 55 &  9532127 & 2128111975407524480 & 15.072 & 0.873 & 5808 & 1.049 & G5 & 10.19 & 0.3753 & 15 & Y \\
 56 &  9592579 & 2128072564788836096 & 15.049 & 0.873 & 5807 & 1.048 & G5 & 10.44 & 0.6905 & 15 & Y \\
 57 &  9470987 & 2128099670319729408 & 15.169 & 0.880 & 5786 & 1.044 & G5 &  9.68 & 0.1934 & 15 & Y \\
 58 &  9655315 & 2128120840220460544 & 15.101 & 0.883 & 5777 & 1.042 & G5 & 10.67 & 0.3633 & 15 & Y \\
 59 &  9716650 & 2128134549756166272 & 15.190 & 0.885 & 5771 & 1.040 & G5 & 10.97 & 0.4127 & 15 & Y \\
 60 &  9715987 & 2128122283329454208 & 15.141 & 0.887 & 5765 & 1.039 & G5 & 10.53 & 0.2703 & 15 & Y \\
 61 &  9655276 & 2128120702781491072 & 15.049 & 0.888 & 5763 & 1.038 & G5 & 10.15 & 0.2139 & 15 & Y \\
 62 &  9716502 & 2128132694330246144 & 15.124 & 0.890 & 5757 & 1.036 & G5 &  9.69 & 0.3416 & 15 & Y \\
 63 &  9776546 & 2128148018773587200 & 15.127 & 0.892 & 5751 & 1.034 & G5 & 10.58 & 0.5511 & 15 & Y \\
 64 &  9836149 & 2128198836827103616 & 15.147 & 0.900 & 5725 & 1.027 & G5 & 10.41 & 0.3628 & 15 & Y \\
 65 &  9896700 & 2128152760418078592 & 15.181 & 0.905 & 5710 & 1.022 & G5 & 10.43 & 0.2606 & 15 & Y \\
 66 &  9716955 & 2128133518964049920 & 15.260 & 0.906 & 5707 & 1.021 & G5 & 11.03 & 0.4030 & 15 & Y \\
 67 & 10018969 & 2128527487723005184 & 15.278 & 0.907 & 5704 & 1.020 & G5 & 11.48 & 0.5578 & 15 & Y \\
 68 &  9777828 & 2128138054449629824 & 15.277 & 0.913 & 5686 & 1.015 & G5 & 11.24 & 0.3175 & 15 & Y \\
 69 &  9838036 & 2128142933532501376 & 15.261 & 0.916 & 5675 & 1.012 & G5 & 13.26 & 0.7145 & 13 & S \\
 70 &  9095289 & 2127962819783332608 & 15.287 & 0.917 & 5673 & 1.011 & G5 & 13.56 & 0.8035 & 15 & S \\
 71 &  9718106 & 2080095615429775360 & 15.325 & 0.922 & 5657 & 1.006 & G8 & 11.11 & 0.2655 & 15 & Y \\
 72 &  9593885 & 2128119087873695232 & 15.201 & 0.944 & 5589 & 0.986 & G8 & 10.22 & 0.4332 & 15 & Y \\
 73 &  9655172 & 2128120256104862464 & 15.413 & 0.948 & 5578 & 0.983 & G8 & 11.11 & 0.2305 & 15 & Y \\
 74 &  9838215 & 2128139600637881472 & 15.575 & 0.966 & 5524 & 0.969 & G8 & 11.71 & 0.4624 & 15 & Y \\
 75 &  9531467 & 2128019994389145728 & 15.510 & 0.974 & 5501 & 0.963 & G8 & 11.49 & 0.3041 & 15 & Y \\
 76 &  9957187 & 2128158601573673728 & 15.414 & 0.978 & 5486 & 0.959 & G8 & 10.93 & 0.3486 & 15 & Y \\
 77 &  9776909 & 2128146713103563264 & 15.531 & 0.981 & 5480 & 0.958 & G8 & 11.84 & 0.3286 & 12 & Y \\
 78 &  9656555 & 2080096886739921152 & 15.588 & 0.982 & 5477 & 0.957 & G8 & 15.08 & $\cdots$ &  1 & S \\
 79 &  9287620 & 2079958966750150272 & 15.632 & 0.985 & 5465 & 0.954 & G8 & 12.17 & $\cdots$ &  1 & Y \\
 80 &  9469735 & 2128018379481370880 & 15.585 & 0.986 & 5464 & 0.954 & G8 &  9.62 & 0.3344 & 15 & Y \\
 81 &  9471316 & 2128098746908279680 & 15.561 & 0.989 & 5454 & 0.951 & G8 & 11.80 & 0.5301 & 15 & Y \\
 82 &  9777281 & 2128140253472804736 & 15.559 & 0.990 & 5451 & 0.950 & G8 & 11.26 & 0.3377 & 14 & Y \\
 83 &  9776413 & 2128146335146361856 & 15.495 & 0.999 & 5425 & 0.944 & G8 &  9.88 & 0.2627 & 15 & Y \\
 84 &  9655609 & 2128108024038073216 & 15.656 & 1.001 & 5420 & 0.943 & G8 & 11.02 & 0.4898 & 15 & Y \\
 85 &  9656371 & 2080090667627204736 & 15.668 & 1.007 & 5402 & 0.938 & K0 & 10.90 & 0.3827 & 15 & Y \\
 86 &  9654627 & 2128122420768295808 & 15.731 & 1.009 & 5396 & 0.937 & K0 & 11.11 & 0.2504 & 14 & Y \\
 87 &  9595006 & 2128128055765212416 & 15.735 & 1.014 & 5382 & 0.933 & K0 & 11.02 & 0.3848 & 15 & Y \\
 88 &  9469908 & 2128020166187804288 & 15.800 & 1.028 & 5343 & 0.923 & K0 & 10.98 & 0.1800 & 14 & Y \\
 89 &  9469760 & 2128018826157984512 & 15.762 & 1.033 & 5330 & 0.920 & K0 & 11.08 & 0.5434 & 15 & Y \\
 90 &  9716177 & 2128121218177579392 & 15.966 & 1.065 & 5241 & 0.898 & K0 & 11.11 & 0.3808 & 14 & Y \\
 91 &  9838358 & 2128514053065223808 & 15.942 & 1.066 & 5238 & 0.897 & K0 & 11.11 & 0.5126 & 13 & Y \\
 92 &  9716378 & 2128145304354240128 & 16.004 & 1.082 & 5193 & 0.887 & K0 & 11.34 & 0.5629 & 13 & Y \\
 93 &  9655435 & 2128132586950838912 & 15.989 & 1.086 & 5185 & 0.885 & K0 & 11.98 & 0.3800 & 14 & Y \\
 94 &  9654919 & 2128119671989269504 & 15.997 & 1.089 & 5175 & 0.883 & K0 & 11.11 & 0.5609 & 14 & Y \\
 95 &  9654970 & 2128121561774914304 & 16.111 & 1.099 & 5151 & 0.877 & K0 & 10.46 & 0.2355 & 15 & Y \\
 96 &  9837012 & 2128148671608654720 & 16.163 & 1.102 & 5141 & 0.875 & K0 & 10.66 & 0.1761 & 14 & Y \\
 97 &  9286138 & 2128006662810036224 & 16.092 & 1.103 & 5141 & 0.875 & K0 & 10.87 & 0.7441 & 15 & Y \\
 98 &  9655225 & 2128121012019139584 & 16.036 & 1.103 & 5140 & 0.875 & K0 & 11.11 & 0.6488 & 15 & Y \\
 99 &  9655469 & 2128108814312039040 & 16.068 & 1.108 & 5126 & 0.872 & K0 & 11.03 & 0.7185 & 15 & Y \\
100 &  9594099 & 2128118916075039360 & 16.139 & 1.109 & 5123 & 0.871 & K0 & 10.17 & 0.3487 & 14 & Y \\
101 &  9835687 & 2128178976898255744 & 16.136 & 1.111 & 5119 & 0.870 & K0 & 10.67 & 0.0330 &  3 & Y \\
102 &  9775249 & 2128175334765924608 & 16.004 & 1.113 & 5114 & 0.869 & K0 & 10.00 & 0.3087 & 10 & Y \\
103 &  9655677 & 2128131418719632000 & 16.148 & 1.123 & 5089 & 0.863 & K0 & 10.71 & 0.6440 & 15 & Y \\
104 &  9654808 & 2128123069303321472 & 16.033 & 1.125 & 5083 & 0.862 & K0 & 11.15 & 0.2196 & 13 & Y \\
105 &  9836986 & 2128148568529429120 & 16.241 & 1.129 & 5075 & 0.860 & K0 & 10.94 & 0.5297 & 14 & Y \\
106 &  9777063 & 2128135408749666304 & 16.295 & 1.139 & 5048 & 0.854 & K2 & 11.11 & 0.5649 & 13 & Y \\
107 &  9594718 & 2128104141387273344 & 16.240 & 1.144 & 5038 & 0.852 & K2 & 13.16 & 0.7599 & 13 & S \\
108 &  9532052 & 2128112181565948800 & 16.172 & 1.149 & 5024 & 0.848 & K2 & 10.44 & 0.2166 & 11 & Y \\
109 &  9716694 & 2128134446676949120 & 16.027 & 1.160 & 4998 & 0.843 & K2 & 10.37 & 0.6846 & 14 & Y \\
110 &  9715637 & 2128124173114998272 & 16.106 & 1.161 & 4997 & 0.842 & K2 & 10.90 & 0.5701 &  7 & Y \\
111 &  9531975 & 2128111700529584000 & 16.293 & 1.170 & 4975 & 0.837 & K2 & 10.99 & 0.1751 &  7 & Y \\
112 &  9776327 & 2128149358803331840 & 16.245 & 1.171 & 4974 & 0.837 & K2 & 10.78 & 0.1928 & 15 & Y \\
113 &  9717373 & 2128135649267419904 & 16.425 & 1.190 & 4929 & 0.827 & K2 & 11.18 & 0.2274 & 14 & Y \\
114 &  9716302 & 2128144445360737408 & 16.468 & 1.225 & 4852 & 0.809 & K2 & 10.39 & 0.1642 &  7 & Y \\
115 &  9531969 & 2128111700529585024 & 16.526 & 1.225 & 4851 & 0.809 & K2 & 11.20 & 0.4061 & 11 & Y \\
116 &  9595731 & 2080089877353291136 & 16.521 & 1.230 & 4842 & 0.807 & K2 & 11.85 & 0.5147 & 10 & Y \\
117 &  9896609 & 2128158910811242496 & 16.593 & 1.247 & 4806 & 0.799 & K2 & 11.28 & 0.3487 & 15 & Y \\
118 &  9593995 & 2128118469398421632 & 16.638 & 1.253 & 4793 & 0.796 & K2 & 11.29 & 0.1830 & 11 & Y \\
119 &  9717161 & 2128136714419757952 & 16.694 & 1.262 & 4775 & 0.792 & K2 & 13.11 & 0.5917 & 12 & S \\
120 &  9716707 & 2128134618476431616 & 16.902 & 1.302 & 4698 & 0.774 & K4 & 10.78 & 0.3101 & 11 & Y \\
121 &  9838166 & 2128139497558660224 & 16.751 & 1.307 & 4688 & 0.772 & K4 & 12.16 & 0.2646 &  7 & Y \\
122 &  9657183 & 2080094000522047872 & 16.801 & 1.308 & 4686 & 0.772 & K4 & 10.78 & 0.0073 &  2 & Y \\
123 &  9655556 & 2128132488171813376 & 16.770 & 1.311 & 4681 & 0.771 & K4 & 13.16 & 0.5852 & 11 & S \\
124 &  9594399 & 2128108466414347264 & 16.840 & 1.323 & 4659 & 0.766 & K4 & 11.79 & 0.4487 & 11 & Y \\
125 &  9594079 & 2128118572477645824 & 16.813 & 1.327 & 4652 & 0.764 & K4 & 12.27 & 0.4963 & 12 & Y \\
126 &  9838228 & 2128139634997621376 & 16.863 & 1.341 & 4626 & 0.758 & K4 & 12.02 & 0.1978 &  6 & Y \\
127 &  9897654 & 2128144204842804992 & 16.904 & 1.347 & 4616 & 0.756 & K4 & 11.02 & 0.0965 &  6 & Y \\
128 &  9593626 & 2128116540951535744 & 16.871 & 1.362 & 4590 & 0.750 & K4 & 11.61 & 0.1967 & 11 & Y \\
129 &  9716918 & 2128133518964044032 & 16.910 & 1.366 & 4583 & 0.749 & K4 & 11.92 & 0.8190 & 11 & Y \\
130 &  9593890 & 2128119087873694208 & 16.760 & 1.369 & 4578 & 0.748 & K4 & 11.53 & 0.1579 & 11 & Y \\
131 &  9895567 & 2128179114337220608 & 17.001 & 1.383 & 4555 & 0.743 & K4 & 12.43 & 0.3667 &  4 & Y \\
132 &  9897050 & 2128155161299953024 & 16.975 & 1.390 & 4543 & 0.740 & K4 & 12.10 & 0.2839 &  6 & Y \\
133 &  9596065 & 2080085685465236864 & 16.908 & 1.400 & 4527 & 0.737 & K4 & 12.30 & 0.1932 &  2 & Y \\
134 &  9533430 & 2080064554225921792 & 17.054 & 1.407 & 4516 & 0.735 & K4 & 11.91 & 0.2322 &  6 & Y \\
135 &  9897078 & 2128155096879780736 & 17.033 & 1.413 & 4507 & 0.733 & K4 & 12.35 & 0.2670 &  6 & Y \\
136 &  9838655 & 2080477558281297152 & 17.138 & 1.418 & 4499 & 0.731 & K4 & 12.54 & 0.0794 &  2 & Y \\
137 &  9656681 & 2080095993386737408 & 17.183 & 1.449 & 4451 & 0.721 & K5 & 12.21 & 0.1257 &  5 & Y \\
138 &  9716678 & 2128134442376853120 & 17.334 & 1.484 & 4400 & 0.711 & K5 & 18.83 & 0.4072 & 10 & S \\
139 &  9716095 & 2128122004151549184 & 17.353 & 1.491 & 4391 & 0.709 & K5 & 12.93 & 0.9576 & 10 & Y \\
140 &  9471356 & 2128098884347245184 & 17.257 & 1.492 & 4390 & 0.708 & K5 & 12.64 & 0.3280 &  6 & Y \\
141 &  9593970 & 2128118228880232704 & 17.340 & 1.492 & 4389 & 0.708 & K5 & 17.68 & 0.8943 &  4 & S \\
142 &  9471158 & 2128098433369035392 & 17.394 & 1.492 & 4389 & 0.708 & K5 & 13.22 & 0.0954 &  3 & Y \\
143 &  9655970 & 2128130387927588992 & 17.322 & 1.506 & 4370 & 0.704 & K5 & 12.74 & 0.3492 & 11 & Y \\
144 &  9594642 & 2128107165044245888 & 17.297 & 1.507 & 4368 & 0.704 & K5 & 11.55 & 0.2238 & 11 & Y \\
145 &  9655566 & 2128132282013368960 & 17.119 & 1.511 & 4364 & 0.703 & K5 & 11.26 & 0.6347 &  6 & Y \\
146 &  9655067 & 2128120492322841728 & 17.404 & 1.512 & 4361 & 0.702 & K5 & 12.95 & 0.1459 &  8 & Y \\
147 &  9956996 & 2128208144016975488 & 17.313 & 1.517 & 4355 & 0.701 & K5 & 12.76 & $\cdots$ &  1 & Y \\
148 &  9470057 & 2128019307193856768 & 17.125 & 1.518 & 4353 & 0.701 & K5 & 12.88 & 0.0759 &  2 & Y \\
149 &  9655054 & 2128121389977163904 & 17.378 & 1.528 & 4340 & 0.698 & K5 & 13.66 & 0.5414 &  6 & Y \\
150 &  9897266 & 2128156673128579328 & 17.483 & 1.555 & 4305 & 0.689 & K5 & 12.99 & 0.2186 &  5 & Y \\
151 &  9530907 & 2128066998511354880 & 17.500 & 1.565 & 4292 & 0.686 & K5 & 12.96 & $\cdots$ &  1 & Y \\
152 &  9594284 & 2128108436354871936 & 17.565 & 1.566 & 4291 & 0.686 & K5 & 13.32 & 0.4442 &  8 & Y \\
153 &  9471304 & 2128098712548536320 & 17.398 & 1.572 & 4284 & 0.684 & K5 & 10.87 & $\cdots$ &  1 & R \\
154 &  9836401 & 2128197904814375808 & 17.471 & 1.579 & 4275 & 0.682 & K5 & 12.79 & 0.3478 &  3 & Y \\
155 &  9836918 & 2128151695265618944 & 17.564 & 1.582 & 4271 & 0.681 & K5 & 12.04 & 0.3328 &  5 & Y \\
156 &  9837294 & 2128153619411014400 & 17.558 & 1.585 & 4267 & 0.680 & K5 & 13.16 & 0.3089 &  5 & Y \\
157 &  9776201 & 2128149668040979456 & 17.611 & 1.596 & 4253 & 0.677 & K5 & 12.79 & $\cdots$ &  1 & Y \\
158 &  9593943 & 2128118327657867776 & 17.622 & 1.598 & 4250 & 0.676 & K5 & 13.16 & $\cdots$ &  1 & Y \\
159 &  9656537 & 2080097195977577472 & 17.545 & 1.653 & 4185 & 0.660 & K7 & 13.35 & 0.1459 &  3 & Y \\
160 &  9533567 & 2080088537323743360 & 17.770 & 1.662 & 4175 & 0.658 & K7 & 13.65 & $\cdots$ &  1 & Y \\
161 &  9594614 & 2128107783519544448 & 17.813 & 1.706 & 4124 & 0.646 & K7 & 12.44 & $\cdots$ &  1 & Y \\
162 &  9895707 & 2128202410239878784 & 17.869 & 1.716 & 4114 & 0.643 & K7 & 14.05 & $\cdots$ &  1 & Y \\
163 &  9717490 & 2080097260396182016 & 17.945 & 1.737 & 4090 & 0.637 & K7 & 13.91 & 0.0926 &  2 & Y \\
164 &  9716651 & 2128134511096352640 & 18.135 & 1.768 & 4058 & 0.630 & K7 & 14.43 & $\cdots$ &  1 & Y \\
165 &  9656400 & 2128129632013326976 & 18.035 & 1.770 & 4055 & 0.629 & K7 & 12.23 & 0.3799 &  5 & Y \\
166 &  9716098 & 2128122042811294208 & 18.133 & 1.784 & 4041 & 0.626 & K7 & 14.43 & $\cdots$ &  1 & Y \\
167 &  9595606 & 2080089976131469568 & 18.220 & 1.835 & 3988 & 0.616 & K7 & 14.43 & $\cdots$ &  1 & Y \\
168 &  9836690 & 2128152068923234432 & 18.234 & 1.836 & 3987 & 0.616 & K7 & 14.20 & $\cdots$ &  1 & Y \\
169 &  9654921 & 2128119706349010944 & 18.365 & 1.933 & 3891 & 0.598 & M0 & 14.34 & $\cdots$ &  1 & Y \\
170 &  9655977 & 2128128910458794112 & 18.438 & 1.939 & 3885 & 0.597 & M0 & 15.15 & $\cdots$ &  1 & Y \\
171 &  9655330 & 2128108981810511616 & 18.377 & 2.001 & 3827 & 0.583 & M0 & 15.96 & $\cdots$ &  1 & Y \\
172 &  9837752 & 2128143478993406592 & 18.931 & 2.100 & 3740 & 0.558 & M1 & 10.55 & $\cdots$ &  1 & R \\
\enddata
\tablecomments{The table is sorted according to \gbr, so also approximately by $G$. Effective temperature \teff\ is calculated with an empirical color--temperature relation. \prot is measured for every available Quarter between 2 and 16 with Lomb--Scargle periodograms. The standard deviation is adopted as the uncertainty, $\sigma \prot$. The number of Quarters of data used is $N_{\rm Q}$.
The code column denotes rapid (``R''), slow (``S''), 
and typical stars (``Y'' for yes this is consistent with the cluster sequence).} 
\end{deluxetable*}

\end{document}